# Multizone Modeling of Black Hole Accretion and Feedback in 3D GRMHD: Bridging Vast Spatial and Temporal Scales

Hyerin Cho (조혜린)[1,2], Ben S. Prather[3], Kung-Yi Su[1,2], Ramesh Narayan[1,2], and Priyamvada Natarajan[2,4,5]

[1] Center for Astrophysics | Harvard & Smithsonian, 60 Garden Street, Cambridge, MA 02138, USA
[2] Black Hole Initiative at Harvard University, 20 Garden Street, Cambridge, MA 02138, USA
[3] CCS-2, Los Alamos National Laboratory, P.O. Box 1663, Los Alamos, NM 87545, USA
[4] Department of Astronomy, Yale University, Kline Tower, 266 Whitney Avenue, New Haven, CT 06511, USA
[5] Department of Physics, Yale University, P.O. Box 208121, New Haven, CT 06520, USA

## Abstract

Simulating accretion and feedback from the horizon scale of supermassive black holes (SMBHs) out to galactic scales is challenging because of the vast range of scales involved. Elaborating on H. Cho et al., we describe and test a "multizone" technique, which is designed to tackle this difficult problem in three-dimensional general relativistic magnetohydrodynamic (GRMHD) simulations. While short-timescale variability should be interpreted with caution, the method is demonstrated to be well-suited for finding dynamical steady states over a wide dynamic range. We simulate accretion on a nonspinning SMBH ($a_* = 0$) using initial conditions and the external galactic potential from a large-scale galaxy simulation and achieve a steady state over eight decades in radius. As found in H. Cho et al., the density scales with radius as $\rho \propto r^{-1}$ inside the Bondi radius $R_B$, which is located at $R_B = 2 \times 10^5\, r_g$ ($\approx$60 pc for M87), where $r_g$ is the gravitational radius of the SMBH; the plasma-$\beta$ is $\sim$ unity, indicating an extended magnetically arrested state; the mass accretion rate $\dot{M}$ is $\approx$1% of the analytical Bondi accretion rate $\dot{M}_{\rm B}$; and there is continuous energy feedback out to $\approx 100 R_B$ (or beyond $>$ kpc) at a rate $\approx 0.02 \dot{M} c^2$. Surprisingly, no ordered rotation in the external medium survives as the magnetized gas flows to smaller radii, and the final steady solution is very similar to when the exterior has no rotation. Using the multizone method, we simulate GRMHD accretion over a wide range of Bondi radii, $R_B \sim 10^2 - 10^7\, r_g$, and find that $\dot{M}/\dot{M}_B \approx (R_B/6\, r_g)^{-0.5}$.

## 1. Introduction

In the nearby Universe, nearly all galaxies appear to host central supermassive black holes (SMBHs). Empirical data show that the masses of these SMBHs are correlated to the stellar mass and the velocity dispersion of stars in the inner regions of their host galaxies (e.g., J. Magorrian et al. 1998; L. Ferrarese & D. Merritt 2000; K. Gebhardt et al. 2000; J. Kormendy & L. C. Ho 2013). The derived scaling relations between the black hole (BH) mass $M_\bullet$ and the galaxy stellar mass $M_*$ suggest that the growth of the SMBHs and the assembly of the stellar component are coupled. Although star formation and SMBH growth both consume gas, the relevant processes for each occur on widely disparate spatial and temporal scales. In fact, the details of how gas makes it into the nuclear regions of galaxies on subparsec scales from larger physical scales of even kiloparsecs let alone Mpc is poorly understood at present. Therefore, while it is clear that SMBH feeding and feedback are essential in connecting such disparate scales, how precisely the connection works remains an open question.

The limits of computational power and resolution have made numerically tackling the coupled SMBH feeding and the feedback problem extremely challenging. While cosmological simulations of structure formation do trace large-scale gas flows from kiloparsec scales, they need special prescriptions to model flows near central SMBHs due to lack of resolution (e.g., D. Sijacki et al. 2015; Y. Rosas-Guevara et al. 2016; R. Weinberger et al. 2018; A. Ricarte et al. 2019; Y. Ni et al. 2022; S. Wellons et al. 2023). Explicit "subgrid models" are therefore adopted in simulations to permit gas accretion and inject feedback at approximately the Bondi radius (e.g., Y. Li & G. L. Bryan 2014; D. Fiacconi et al. 2018; D. Anglés-Alcázar et al. 2021; K.-Y. Su et al. 2021; R. Y. Talbot et al. 2021; S. Koudmani et al. 2024; D. Rennehan et al. 2024; R. Weinberger et al. 2023). Meanwhile, feedback from accreting SMBHs is required in cosmological simulations to solve the "cooling flow problem" and to modulate star formation to reproduce the observed properties of massive galaxies (see, e.g., K.-Y. Su et al. 2019, which excludes other possibilities). In these simulations, typically gas flows are resolved only at the Bondi radius $R_B$ and beyond, and any gas particles that cross $R_B$ are assumed to be accreted by the SMBH (see, e.g., the SMBH feeding implementation in the Romulus suite, M. Tremmel et al. 2017; and in the ASTRID simulation, Y. Ni et al. 2022).

Independently, smaller scales near the SMBH have been tackled with a different computational approach, using general relativistic magnetohydrodynamic (GRMHD) simulations that self-consistently trace the dynamics of the gas and the magnetic fields near the BH, essentially from first principles (e.g., S. S. Komissarov 1999; C. F. Gammie et al. 2003; O. Porth et al. 2019; R. Narayan et al. 2022; K. Chatterjee et al. 2023). However, these simulations typically achieve a steady state only out to $\sim 10^2\, r_g$, where $r_g$ is the gravitational radius corresponding to the BH mass $M_\bullet$, which is far short of the Bondi radius $R_B > 10^5\, r_g$ (see Equation (1) for the definitions of $r_g$ and $R_B$). Moreover, the idealized settings explored in GRMHD simulations do not take the larger-scale cosmological environment into account. Bridging cosmological scales and near-horizon scales while self-consistently following gas flows fully in GRMHD is the current challenge.

Some recent efforts to bridge the vastly different scales in the feeding-feedback problem have used a variety of numerical techniques. These include (i) zoom-in methods that involve nested simulations wherein subsequent smaller-scale runs are initialized from larger-scale simulations (e.g., P. F. Hopkins & E. Quataert 2010; S. M. Ressler et al. 2020; M. Guo et al. 2023, 2024); (ii) using Lagrangian hyperrefinement techniques that permit augmenting resolution on small portions of the simulation box (e.g., D. Anglés-Alcázar et al. 2021; P. F. Hopkins et al. 2024a, 2024b); and (iii) running GRMHD simulations for an extended duration to slowly reach convergence on larger scales (e.g., A. Lalakos et al. 2022, 2024; N. Kaaz et al. 2023). Deployment of these methodologies has yielded some success in studying accretion from large scales onto horizon scales of the central SMBHs. However, due to limitations in two-way communication from the inner regions to the outskirts, feedback from the SMBH has not been followed out to galaxy scales in these previous works.

In H. Cho et al. (2023), we presented the first attempt to track both accretion and feedback from the event horizon to galaxy scales, using a "multizone" computational method that permits simultaneously tracing both gas accretion as well as feedback across 7 orders of magnitude in radius. A key feature of our method is that it directly seeks the global dynamic steady-state solution to the problem without attempting to solve the detailed time-dependent dynamics.

Using the proposed method, we studied purely hydrodynamic H. Bondi's (1952) accretion as well as its magnetized analog. For the classical hydrodynamic spherically symmetric Bondi accretion problem, our results were in excellent agreement with the analytic general relativistic solution. Meanwhile, in the case when the accreting gas is magnetized, the SMBH magnetosphere was found to become saturated with a strong magnetic field, which modified the dynamics over the entire volume inside the Bondi radius. In particular, we reported that (i) the density profile varied as $\sim r^{-1}$ and not as $r^{-3/2}$ as for the purely hydrodynamic Bondi case; (ii) the mass accretion rate $\dot{M}$ on to the BH was suppressed by more than 2 orders of magnitude relative to the Bondi rate $\dot{M}_{\rm B}$; and (iii) there was continuous energy feedback from the vicinity of the SMBH to the external medium at a level of $\sim 10^{-2}\dot{M}c^2$. We found that energy is transported via turbulent convection, which is triggered by magnetic reconnection near the SMBH and it is this feedback mechanism that couples these disparate scales. Thus, physical processes on small scales, associated with strong magnetic fields that accumulate near the horizon, were shown for the first time to impact the dynamics of gas flow far from the SMBH. Notably, these GRMHD simulations considered nonspinning BHs, so the feedback mechanism is quasi-isotropic and different from the directed feedback associated, for example, with collimated jets from spinning accreting BHs.

In the present paper, we provide the full implementation details of our multizone methodology, expanding on the concise description in H. Cho et al. (2023). We show tests of the validity of the method and demonstrate that it is able to faithfully reproduce dynamical steady states extending over many decades in radius, and we also present several new simulations and results.

While spherical hydrodynamical accretion onto a point mass is fully understood, the addition of magnetic fields makes the problem much harder. As already noted by V. F. Shvartsman (1971), frozen-in magnetic field lines in the accreting plasma cause the magnetic energy density to increase steeply with decreasing radius, as $r^{-4}$, as gas flows in, thereby quickly overwhelming the gravitational and thermal energy of the gas. For accretion to continue, the field has to reconnect and dissipate some of its energy. I. V. Igumenshchev & R. Narayan (2002) carried out numerical MHD simulations of magnetized spherical accretion and showed that the dissipated energy flows outward via a form of convection. Moreover, the mass accretion rate is drastically reduced compared to the hydrodynamic Bondi case. These results were confirmed soon thereafter by U.-L. Pen et al. (2003), who refer to the resulting flow as "magnetically frustrated convection." They noted that the density of the spherically accreting magnetized gas varies with radius as $\rho \propto r^{-0.8}$, which is very different from the $\rho \propto r^{-3/2}$ prediction of the analytical hydrodynamic Bondi solution.

In related work, I. V. Igumenshchev et al. (2003) simulated the accretion of gas with angular momentum and a frozen-in poloidal magnetic field in GRMHD simulations. They showed that, in the quasi-steady state, the flow becomes highly nonaxisymmetric, with accretion proceeding primarily via narrow streams. R. Narayan et al. (2003) referred to this kind of accretion as a magnetically arrested disk (MAD) and argued that especially for accretion from the central regions of a galaxy onto an SMBH, MAD accretion should occur frequently. The importance of MAD accretion became widely recognized after the work of A. Tchekhovskoy et al. (2011), who showed that the MAD state arises spontaneously in GRMHD simulations of black hole accretion (see also M. Liska et al. 2020), and that MAD accretion onto spinning black holes leads to very powerful jets via the R. D. Blandford & R. L. Znajek (1977, BZ) mechanism.

Results from the Event Horizon Telescope collaboration have demonstrated that MADs offer a good fit to both M87* (Event Horizon Telescope Collaboration et al. 2021) and Sgr A* (Event Horizon Telescope Collaboration et al. 2022) data. It therefore appears that MADs might be common for local SMBHs, most of which appear to accrete at well below the Eddington rate via radiatively inefficient, hot accretion flows (see F. Yuan & R. Narayan 2014 for a review). MAD accretion may therefore account for the majority of SMBHs in the Universe once quasar activity has declined. The rarer, optically luminous quasar population meanwhile might be powered instead by radiatively efficient, thin accretion disks (J. Frank et al. 2002), where the relevant astrophysical processes are very different.

In the present work, we follow up on the initial work reported in H. Cho et al. (2023) and elaborate in greater detail our methodology, along with providing multiple tests of the method as well as results from new applications. We continue to focus on a nonspinning BH, $a_* = 0$, described by the Schwarzschild metric. The adiabatic index of the gas adopted is $\gamma_{\rm ad} = 5/3$. The gravitational radius $r_g$, and the Bondi radius $R_B$ (inside which the gravity from the BH dominates), are given

by:

$$r_g \equiv \frac{GM_\bullet}{c^2}, \qquad R_B \equiv \frac{GM_\bullet}{c_{s,\infty}^2}, \tag{1}$$

where $c_{s,\infty} = \sqrt{\gamma_{\rm ad} T_\infty}$ is the sound speed at large radii well beyond $R_B$. We use units such that lengths are expressed in $r_g$ and times in $t_g \equiv r_g/c$.

The outline of the paper is as follows: In Section 2, we describe the details of our multizone numerical scheme, which is designed to capture two-way communication between the SMBH and the external medium beyond the Bondi radius. In Section 3, we present tests of the multizone method using a simulation with a smaller Bondi radius $R_B \approx 400\, r_g$, where it is possible to cross-compare with results from traditional, nonmultizone simulations. We then proceed to perform multizone simulations with a realistic $R_B \approx 2 \times 10^5\, r_g$ and present new simulations, which include the external gravitational potential of the galaxy and are initialized with realistic outer conditions of the gas and magnetic field from galaxy-scale simulations. These simulations are discussed for the pure hydrodynamic (HD) problem and the more realistic magnetohydrodynamic (MHD) problem in Sections 4 and 5, respectively. In Section 6, we conduct multizone GRMHD simulations for a wide range of Bondi radii, $R_B = 10^2\text{–}10^7\, r_g$, and present the derived scaling relation between the BH mass accretion rate $\dot{M}$ and the Bondi radius. In Section 7, we consider the effect of including nearly Keplerian rotation in the external gas and show, surprisingly, that the results are almost indistinguishable from the no-rotation case. We summarize and conclude in Section 8.

## 2. Numerical Methods

To simulate accretion and feedback accurately all the way to the event horizon, we solve the ideal GRMHD equations. In particular, we conserve the mass flux, $\rho u^\mu$, and the stress–energy tensor,

$$T^{\mu\nu}_{\rm MHD} = (\rho + u + p_g + b^2) u^\mu u^\nu + \left(p_g + \frac{b^2}{2}\right) g^{\mu\nu} - b^\mu b^\nu, \tag{2}$$

along with the ideal MHD induction equation $F^{*\mu\nu}{}_{;\nu} = 0$ (A. M. Anile 1989; S. S. Komissarov 1999; C. F. Gammie et al. 2003). Above, $\rho$ is the rest-mass density; $u$ is the internal energy density; $u^\mu$ is the fluid four-velocity; $b^\mu$ is the magnetic four-vector in the fluid frame; $p_g = (\gamma_{\rm ad} - 1)u$ is the gas pressure; $g^{\mu\nu}$ is the metric; and $F^{*\mu\nu}$ is the dual of the electromagnetic field tensor. The temperature $T$ is defined in relativistic units such that $p_g = \rho T$. A conservative shock-capturing scheme is used, ensuring local conservation of the stress–energy tensor. It is important to note, however, that the multizone method does not conserve these components at its internal boundaries, allowing for the introduction and removal of material and energy via the boundary conditions, as described in Section 2.2.

We employ the code KHARMA[6] (an acronym for "Kokkos-based High-Accuracy Relativistic Magnetohydrodynamics with Adaptive mesh refinement"), a new open-source code for solving the GRMHD equations in stationary spacetimes, primarily targeting accretion problems. KHARMA is based on the High-Accuracy Relativistic Magnetohydrodynamics (HARM) scheme (C. F. Gammie et al. 2003) but is written from scratch in C++, leveraging the Parthenon framework (P. Grete et al. 2023) and Kokkos programming model (C. Trott et al. 2021) in order to run efficiently on CPUs and GPUs with the same source code. In addition to being fast, KHARMA is written to be readable, modular, and extensible, separating functionality into "packages," representing, e.g., algorithmic components or physics extensions.

We employ the fifth-order WENO reconstruction and a Courant factor of 0.9 for HD and 0.5 for MHD runs. The cell-centered version of flux-interpolated constrained transport (flux-CT in G. Toth 2000) is used to evolve magnetic fields while preserving the divergence $\nabla \cdot B = 0$. The initial conditions, numerical floors, and coordinate systems are different for each run and are described for each case in the following sections. Generally, we employ spherical coordinates with a base resolution of $64^3$ per annulus. Due to the large-scale coherent field structure of the MAD, this is sufficient to resolve the steady-state accretion, and doubling the resolution in each dimension ($128^3$) yielded substantially similar results (see Appendix F, also H. Cho et al. 2023). C. J. White et al. (2019) and L. D. S. Salas et al. (2024) have demonstrated that even higher resolutions produce a similar accretion flow structure as the runs corresponding to our $128^3$ resolution simulations.

### 2.1. Multizone Method

The challenge facing any attempt to simulate accretion and feedback in galactic nuclei is the vast range of length scales and timescales involved in the computation. We are interested in following the evolution of the system on length scales (much) larger than the Bondi radius, $R_B \gtrsim 10^5\, r_g$, i.e., on timescales (much) longer than $t_B \equiv (R_B/r_g)^{3/2}\, t_g \approx 10^8\, t_g$. The accretion and feedback physics, however, are dominated by energetic processes near the BH, which occur on timescales $\approx t_g$. Given our desire to spatially resolve the details of the accretion flow near the horizon, e.g., to resolve the BZ process adequately, the smallest grid cells will typically have a size $\lesssim 0.01\, r_g$, which means that the Courant-limited time step in a numerical simulation is $\Delta t \approx 0.01\, t_g$. To run a simulation with such a small time step for a total duration of $\gtrsim 10\, t_B$ (the minimum for obtaining an approximate steady state up to $R_B$) will require $\gtrsim 10^{11}$ time steps. This is impractical.

We have developed an approximate method to tackle this challenging problem. A bird's eye view of the method was described in H. Cho et al. (2023), and we provide more details here. Our simulation volume extends from an inner radius $r_{\rm in} = r_g$ to an outer radius that lies well outside the Bondi radius, $r_{\rm out} \gg R_B$. In the largest simulation described in this paper, we span $r_{\rm out} \approx 10^8\, r_g$, which corresponds to 30 kpc for a system like M87, and it could be made larger at only a modest computational cost if needed.

Our scheme for partitioning the computational volume is described schematically in Figure 1. We split the large simulation volume from $r_{\rm in}$ to $r_{\rm out}$ into logarithmically evenly spaced spherical annuli or "zones," each covering a limited range of radii. At any given time only one of these annuli is simulated, and the remaining volume of the system is kept frozen. Since the physics in each zone involves only a limited range of characteristic timescales, simulating a single annulus in isolation is tractable. This solves the time-stepping challenge described above and is the primary innovation in our method.

[6] https://github.com/AFD-Illinois/kharma

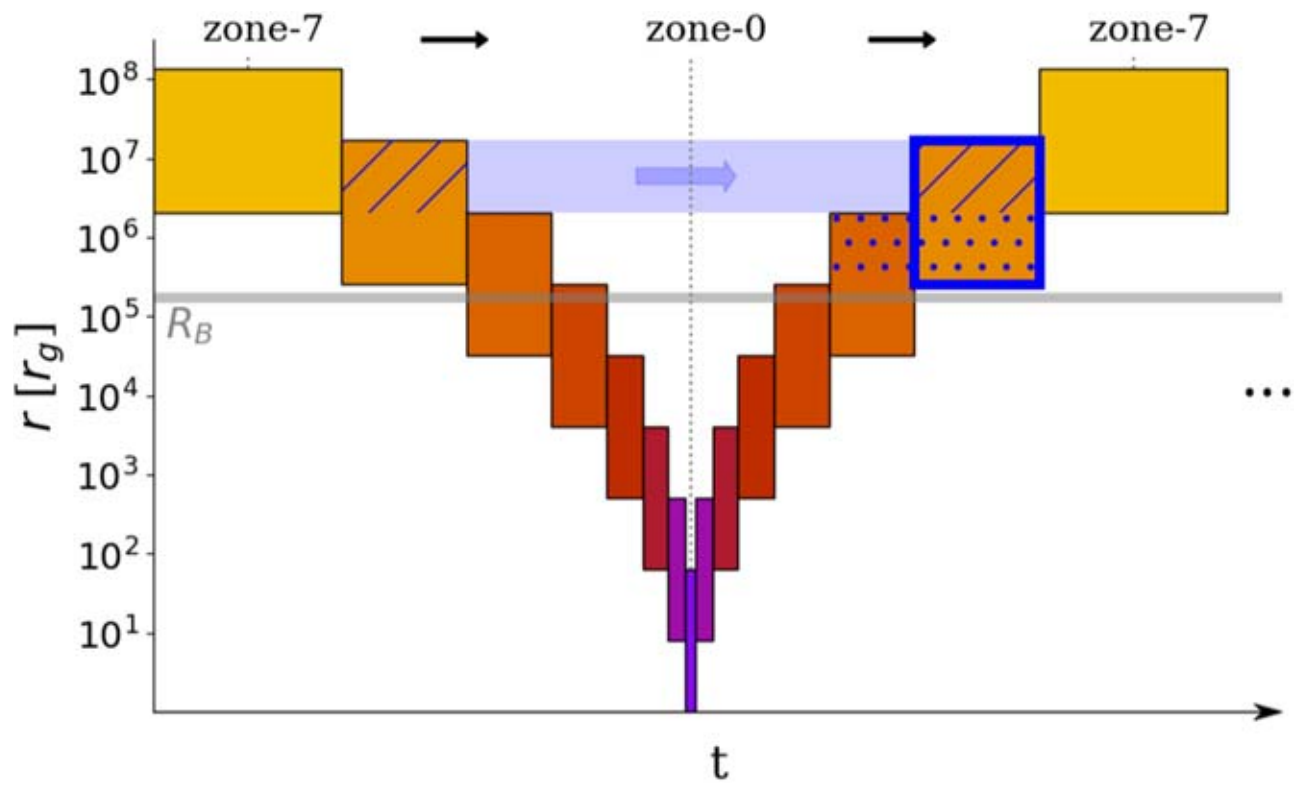

**Figure 1.** Schematics of the multizone method: The different colors represent the different zones being simulated. Radii are shown along the $y$-axis, with zone $i$ extending from an inner radius of $8^i\ r_g$ to an outer radius of $8^{i+2}\ r_g$, where $r_g = GM_\bullet/c^2$ is the gravitational radius. Runtime is shown along the $x$-axis (not to scale). The plot here corresponds to one "V-cycle," advancing the entire domain forward by some time. A complete simulation consists of hundreds of V-cycles to allow full information exchange between the smallest and largest scales.

The computations progress in a "V-cycle," where the active zone is moved in and out a large number of times (named for its passing resemblance to the V-cycle used in multigrid solvers). Although during a given visit, a zone evolves only for a short period of time, integrated over all the visits during a complete simulation, the zone evolves in total for many local characteristic times. This enables the system to achieve a quasi-steady state in all the zones. In addition, as each zone half-overlaps the next, each zone has ample opportunity to exchange information with its neighboring zones, which is critical to maintaining two-way communication between the BH and the host galaxy. Implementation details are described in the following subsections.

#### 2.1.1. Zone Setup

With a total of $n$ zones, the zeroth zone is the innermost annulus and the $(n-1)$th zone is the outermost annulus. The $i$th zone has an inner radius of $r_{i,\mathrm{in}} = 8^i\ r_g$, and an outer radius of $r_{i,\mathrm{out}} = 8^{i+2}\ r_g$. The radial ranges of the various zones are listed in Table 1. Note that the overlap between an annulus $i$ and each of its two neighboring annuli, $i-1$ and $i+1$, is a full 50% of the range of annulus $i$ as measured by $\log r$ (see Figure 1). This large overlap maximizes communication between zones.

The above description uses base 8 to organize the zones. We have found this to be convenient and have used it for all simulations presented in this paper. In Appendix F, we investigate the sensitivity of the results to the choice of base $b$. We show that there is only a minor effect as long as $b \geqslant 8$.

#### 2.1.2. Switching Between Zones

The operation of our multizone method is shown schematically in Figure 1. To start, the outermost $(n-1)$th zone is initialized and simulated for a time $t_{n-1}$, where $t_i$ is the zone-specific runtime for the $i$th zone. The choice of $t_i$ is explained in Section 2.3. Once computation in the $(n-1)$th zone has been completed, the next inner zone $(n-2)$ is simulated for a time $t_{n-2}$. This process of switching the active zone to an inner annulus is repeated until zone 0 is reached, wherein the gas finally reaches the BH horizon. In this sequence of runs, zone 0 is the only zone that is in contact with the BH horizon. After the smallest annulus has been run for a time $t_0$, we then switch direction and march the active zone outwards until the outermost $(n-1)$th zone is reached again. This completes one V-cycle. The cycle is repeated hundreds of times until the simulation reaches a statistical steady state on all scales. In short, we start outside–in, then march out then back in iteratively through the multizone structure to finish the computation.

**Table 1**
Multizone Setup for $n = 8$ Zones

| Zone # | $r_{\mathrm{in}}$ $(r_g)$ | $r_{\mathrm{out}}$ $(r_g)$ |
|---|---|---|
| 0 | $8^0 = 1$ | $8^2$ |
| 1 | $8^1$ | $8^3$ |
| ⋮ | ⋮ | ⋮ |
| $i$ | $8^i$ | $8^{(i+2)}$ |
| ⋮ | ⋮ | ⋮ |
| 7 | $8^7$ | $8^9 \approx 1.3 \times 10^8$ |

The simulation is globally initialized with a state that can be very far from the final steady-state configuration (e.g., constant density). The initialization of each active zone is done as follows. During the first half of the very first V-cycle, if active zone $i$ has a larger neighboring zone $(i+1)$, which was the previously active zone, then the overlapping range of radii $[r_{i+1,\mathrm{in}}, r_{i,\mathrm{out}}]$ is initialized by copying data from the last dump of zone $(i+1)$. The remaining half $[r_{i,\mathrm{in}}, r_{i+1,\mathrm{in}})$ is filled with the global initial conditions.

The above only applies to the first half of the very first V-cycle. Starting with the second half of the first V-cycle and in all subsequent V-cycles, half of the active zone $i$ is initialized with data from the just-previously active zone (either $i-1$ or $i+1$) over the volume where the two annuli overlap. For the remaining logarithmic half, data are copied from the last output of zone $i$ itself. As an example, zone 6 during the outward-moving leg of the V-cycle is highlighted in a blue box in Figure 1. The dotted region overlaps with the just-previously active zone 5, so the last output of zone 5 is copied in the dotted range. The remaining hatched region is then initialized with the last output of zone 6, which was saved during the inward leg of the V-cycle.

### 2.2. Boundary Conditions

The multizone scheme involves splitting the simulation into annuli so there are extra internal radial boundaries compared to typical GRMHD simulations. Here we describe how these internal radial boundaries are handled. At the innermost zone's inner boundary $r_{0,\mathrm{in}}$, the gas flows inward toward the singularity, and this is treated exactly as in standard GRMHD simulations (outflow boundary condition). All other radial boundaries except this innermost boundary follow Dirichlet boundary conditions, where $\rho$, $u$, $u^\mu$, and $b^\mu$ are held fixed to their initial values over the duration of the active zone's runtime. Depending on the direction of the radial velocity $u^r(\theta, \varphi)$ frozen at the boundaries, gas can flow onto or off of the grid through the annulus radial boundaries. Locally, this breaks the conservation of particle number, energy, and momenta of the scheme, by introducing/removing material to the running annulus, which does not necessarily reflect the outflow/inflow from the adjacent annulus. However, good conservation is recovered in the steady state as shown in Figure 2: the profiles of both accretion rate in panel (a) and efficiency in panel (f) are flat for the multizone runs, indicating that when the steady state is reached, the average outflow rate on one side of each boundary matches the inflow rate of the other side.

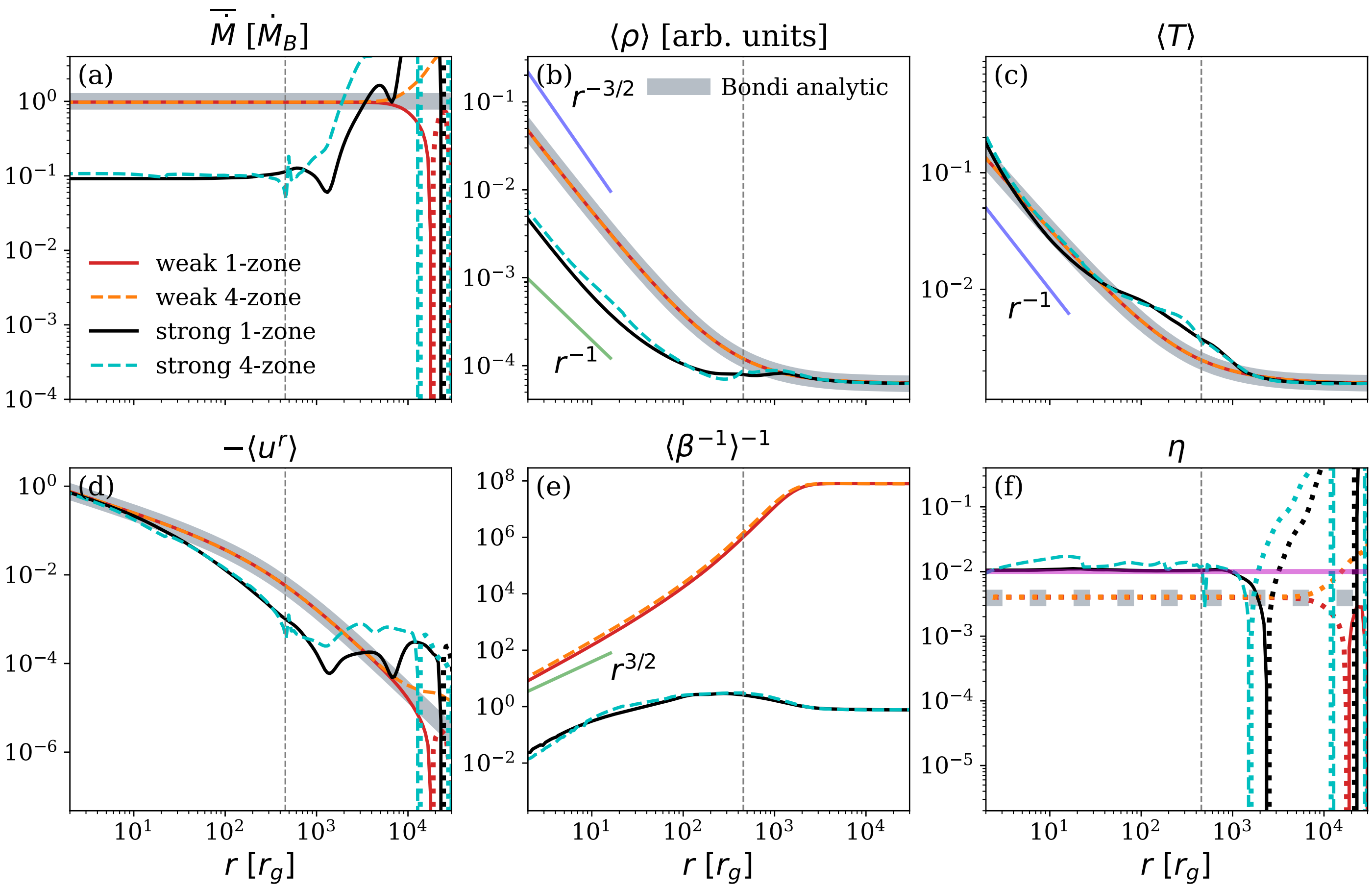

**Figure 2.** Radial profiles corresponding to four simulations of magnetized Bondi accretion (see legend in panel (a)) with a relatively small Bondi radius, $R_B \approx 400\ r_g$. Red solid lines correspond to a run that is initialized with a weak magnetic field ($\beta \approx 10^8$, Section 3.1) and simulated on a single zone, as in usual GRMHD simulations. Dashed orange lines correspond to the exact same problem simulated using our multizone technique, where we divide the simulation domain into multiple annuli and switch between them. Black solid lines (one-zone) and cyan dashed lines (four-zone) correspond to another pair of simulations that are initialized with a strong magnetic field ($\beta \approx 1$, Section 3.2). In both simulation pairs, the one-zone and four-zone runs give closely similar results. The analytical Bondi HD solution is shown as thick gray lines and the Bondi radius is shown as a gray dashed vertical line. The two weak-field simulations closely follow the Bondi HD solution since the magnetic field is too weak to affect the dynamics. The two strong-field simulations deviate significantly from the Bondi solution. For all curves, negative values are shown dotted. The different panels correspond to the following: (a) $t$-averaged accretion rate $\overline{\dot{M}}$ in units of the analytical Bondi accretion rate $\dot{M}_B$. The two strong-field simulations show a reduced accretion rate of $\sim 10^{-1}\dot{M}_B$. (b) $t$-, $\theta$-, $\phi$-averaged density $\rho$. The strong-field simulations have a shallower slope ($\rho$ roughly $\propto r^{-1}$) compared to the Bondi solution (slope $-3/2$). (c) $t$-, $\theta$-, $\phi$-averaged temperature $T$. (d) $t$-, $\theta$-, $\phi$-averaged radial component of the four-velocity $-u^r$. The weak-field simulations agree perfectly with the Bondi solution, but the strong-field simulations deviate significantly. (e) Inverse of the $t$-, $\theta$-, $\phi$-averaged $\beta^{-1}$. For the strong-field runs, $\beta$ is constant around $\sim 1$ across a few orders of magnitude in radius and decreases toward a smaller radius. The radial scaling of $r^{3/2}$ expected for a weak-field simulation is shown by the green line. (f) energy outflow efficiency $\eta$. The weak-field runs show a negative efficiency, $\eta \sim -1/R_B \sim -4 \times 10^{-4}$, consistent with Equation (11). The efficiency switches sign to positive, indicating active outward energy feedback, in the strong-field runs. The feedback efficiency is $\eta \sim 1\%$ (indicated by a thick pink line for reference).

Meanwhile, the magnetic fields are not allowed to cross the radial boundaries under the Dirichlet boundary condition. This is in order to preserve the magnetic field divergence $\nabla \cdot B = 0$ at the annulus boundaries. We refer to this boundary condition for magnetic fields as `bflux0`; see Appendix E.1 for the details of `bflux0` and its effect. Meanwhile reflecting and periodic boundary conditions are used in the $\theta$- and $\varphi$-directions, respectively.

### 2.3. Run Duration for Each Zone

For each zone $i$ we define a characteristic timescale,

$$t_{\mathrm{char},i} = r_{i,\mathrm{out}} \Big/ \sqrt{v_{\mathrm{ff},i}^2 + c_{s,\infty}^2}\,, \tag{3}$$

where $v_{\mathrm{ff},i} = \sqrt{GM_\bullet / r_{i,\mathrm{out}}}$ is the freefall velocity at $r_{i,\mathrm{out}}$. For $r_{i,\mathrm{out}} < R_B$, $t_{\mathrm{char},i}$ is approximately equal to the freefall timescale at the outer edge of the annulus, and for $r_{i,\mathrm{out}} \gg R_B$, it is approximately equal to the sound-crossing time.

The runtime $t_i$ for zone $i$ is set differently for different types of problems. For purely HD problems, it is simply set equal to the characteristic timescale:

$$t_{\mathrm{HD},i} = t_{\mathrm{char},i}. \tag{4}$$

For MHD problems, however, the runtime needs to be chosen with care. If the runtime per zone visit is too long, magnetic tension can accumulate significantly at the radial boundaries (especially the inner boundary) because of the Dirichlet boundary condition on the magnetic field. When a previously active zone's boundary region with its accumulated magnetic stresses moves to the logarithmic center of an active zone, field lines undergo a readjustment that irons out the distortions. However, the readjustment can become very violent if the accumulated stresses are too large, possibly leaving a persistent effect. On the other hand, if the runtime per zone visit is too short, the distortions will have no time to smooth out at all. Since $t_{\mathrm{char},i} \propto r_{i,\mathrm{out}}^{3/2}$ for $r < R_B$, we set the runtime equal to a fraction $8^{-3/2} \approx 0.04$ of $t_{\mathrm{char},i}$, which corresponds to the characteristic time at the logarithmic half radius of the annulus. We include an extra factor of $1/2$ for safety and choose:

$$t_{\mathrm{MHD},\ i} = 0.02\ t_{\mathrm{char},i}. \quad (5)$$

With this prescription for the runtime per zone visit, each active zone is given just enough time to iron out the accumulated field lines from the boundary effect in the previous zone but not given too much time to build up its own boundary effect to unsafe levels. The effect of choosing different fractions is explored in Appendix F and the results are shown to be robust for factors close to the optimal value 0.04.

Finally, the runtime for some MHD problems is capped proportionally to the freefall timescale at the Bondi radius $t_B$

$$t_{\mathrm{MHDcap},i} = 0.02\ \min(t_{\mathrm{char},i}, t_B). \quad (6)$$

This prescription was used for all the MHD runs reported in H. Cho et al. (2023) and is used for the multizone tests (four-zone runs) in Section 3 and small-scale problems in Section 6 of the present paper. The $t_{\mathrm{MHDcap},i}$ type of runtime is useful for comparing multizone results with those obtained from nonmultizone GRMHD simulations (one-zone runs), as explained in Section 3. For all other runs in this paper, we use $t_{\mathrm{MHD},i}$ given in Equation (5).

Finally, since zone 0 does not have a Dirichlet boundary condition at its inner edge but rather has the usual outflow boundary condition inside the BH horizon, its runtime can be as long as needed. We multiply the runtime for zone 0 by an extra factor of 5 (for a total of 0.1 $t_{\mathrm{char},i}$) to ensure that this zone has time to relax fully.

### 2.4. The Power and the Limitations of the Multizone Method

The multizone methodology we have developed is based on the reasonable assumption that the dynamics at each length scale are dominated by processes that operate on a local characteristic time, typically the freefall or sound-crossing time (see Equation (3)). By running each annulus for a similar number of local characteristic times (via the V-cycle shown in Figure 1), all scales approach the local steady state more or less simultaneously. Because information needs to propagate over many orders of magnitude in scale across many annulus boundaries, the method requires hundreds of V-cycles for convergence. Nevertheless, this approach can be much faster than the traditional method of simulating an entire system on a single global grid. In the traditional method, for a problem extending over, say, six decades in radius, the innermost regions will have to be simulated for $10^9$ local $t_{\mathrm{char}}$ before the outermost regions can evolve by a single local $t_{\mathrm{char}}$. This is well-nigh impossible. In the multizone method, all scales are evolved over a few tens of their own local $t_{\mathrm{char}}$.

However, we make note of an important caveat. While the multizone method can generate the approximate steady-state solution for a given system efficiently, it is not designed to answer questions related to time variability, specifically variability on timescales shorter than the timescale of the largest radii in the simulation domain. For information on rapid variability, the traditional simulation method is the appropriate technique.

Although at any given time in the multizone method, only one annulus is active and the rest of the zones are kept frozen, it is important to note that there is constant communication across the radial boundaries of the active annulus. For instance, the radial velocity on the boundaries is nonzero; hence there is a nonzero mass flux $\rho u^r$ on each boundary cell and a nonzero net mass accretion/ejection rate $\dot{M}$ across each wall. Especially at early times, the individual $\dot{M}$ values across the two boundaries of an annulus could be very different and the annulus may gain or lose mass. However, at late times, when the system is in a quasi-steady state, the two $\dot{M}$ values will be nearly the same and the system will settle down to a constant net mass accretion rate from the largest radii down to the horizon.

In a similar fashion, there is a nonzero radial energy flux $-T^r_t$ and angular momentum flux $T^r_\phi$ across the walls, since all the individual terms contributing to the stress–energy tensor (Equation (2)) are nonzero in the boundary cells. Thus, energy and angular momentum fluxes are freely communicated across the boundaries, and at late times these again reach a steady state over the entire radius range of the simulation.

Our treatment of the boundary condition on the magnetic field, as described in Section 2.2 and Appendix E, does impose an undesirable level of rigidity on the flow, which appears to be unavoidable. By switching from our basic `bflux0` boundary condition to the `bflux-const` boundary condition (details are provided in Appendix E.2), we can account for a mean rotation of the magnetic field configuration at each boundary. This is not particularly helpful for the simulations discussed in this paper (see Section 7), though it might prove to be more useful when there is substantial coherent rotation in the accretion flow (or jet) in scenarios we intend to explore in the future. Regardless, even if we succeed in modeling the mean rotation of the flow, any nonaxisymmetric dynamics arising from turbulent fluctuations are necessarily filtered out at the boundaries by the Dirichlet boundary condition. A possible compensating factor is that each annulus is visited a large number of times. Thus, a single simulation generates multiple realizations of the dynamics in the annulus, and in combination, these realizations will provide an estimate of the fluctuation statistics of the accretion flow in that volume. How closely this estimate resembles the true dynamics (in the absence of artificial boundaries) remains to be seen.

Given the above caveats, we reiterate that the multizone method is designed only for the family of problems that evolve to a quasi-steady or slowly varying state. The method is not expected to give reliable results for systems that evolve to wildly different states over timescales shorter than the Bondi timescale.[7] Therefore, as a general rule, cross validation of the multizone method against the conventional one-zone GRMHD method should be attempted wherever feasible, e.g., with a smaller problem as in Section 3. Furthermore, using a pure GRMHD code without radiative cooling limits us presently to simulating only hot accretion flows. Any cooling-related phenomena that occur over the full duty cycle of active galactic nuclei (AGNs), with episodic cold gas accretion (M. Gaspari et al. 2013; Y. Li & G. L. Bryan 2014), will need to be modeled with an external galaxy simulation code. The galaxy simulation can either provide the exterior boundary condition as done here for our runs reported in Sections 4 and 5 or be directly incorporated as the largest zone of the V-cycle in the multizone simulation. However, we note that cold gas has been detected within the Bondi radius for Sgr A* (E. M. Murchikova et al. 2019), which our current multizone implementation does not take into account.

[7] The intermittent jet activity, for instance, reported in A. Lalakos et al. (2024) and A. Galishnikova et al. (2024) evolves over timescales comparable to or longer than the Bondi timescale $t_B$, so we believe that the multizone method should, in future simulations with spinning BHs, be able to capture aspects of the limit cycle behavior that they report.

Notwithstanding the above caveats, there are potential advantageous applications of the multizone method even to systems without well-defined long-term steady states. One important example is using a precomputed library of quasi-steady-state solutions for various outer conditions, e.g., different Bondi radii $R_B$, BH spins $a_*$, external plasma-$\beta$. Such a library will provide a time-dependent BH accretion/feedback prescription in galaxy simulations, in an analogous fashion to the subgrid prescriptions currently in use in galaxy simulations. The time step of galaxy simulations is large enough for unresolved scales to quickly relax to their steady states. Thus, they can directly employ the GRMHD multizone library results to estimate the BH accretion rate $\dot{M}$ and feedback efficiency $\eta$ as the galaxy evolves. This is possible because the multizone method extends the reliable radial range of GRMHD simulations to unprecedented large distances such that the scales overlap with the typical resolved spatial scales of galaxy simulations. This direct bridging of scales between the two classes of simulations is a new opportunity. Therefore, the multizone method findings tabulated into a library as noted above can be adopted with ease for cosmological simulations in the future in lieu of the currently deployed subgrid prescriptions.

## 3. Testing the Multizone Method with Small-scale Simulations

In H. Cho et al. (2023) we tested the multizone method for hydrodynamic Bondi accretion, for which an analytical relativistic solution is available (F. C. Michel 1972; S. L. Shapiro & S. A. Teukolsky 1983; see Appendix A below). The agreement was excellent. However, that was a relatively straightforward test, as it did not involve magnetic fields.

Real accretion flows in galactic nuclei almost certainly involve magnetized plasma, and it has been known for many decades (see the papers cited in Section 1) that magnetic fields strongly perturb the flow dynamics, as also verified by H. Cho et al. (2023). The strongly magnetized problem can be studied only via numerical simulations since no exact analytical solutions are known. If we wish to check the validity of the multizone method for such problems, the only approach available is to run the same problem using traditional GRMHD methods without multiple zones and check if the results are in concordance. However, such a test cannot be carried out for a realistic problem with Bondi radius $R_B \gtrsim 10^5$ (as appropriate, e.g., for Sgr A*, M87) because the dynamic range is far too large for the traditional method.

Here we test the multizone method using a smaller-scale problem with a smaller Bondi radius, $R_B \approx 400\, r_g$ (set by choosing a sonic radius $r_s = 16\, r_g$ in Equation (A3), which corresponds to a very hot ambient medium with $T_\infty \approx 10^{10}$ K). Since the scale separation between the BH horizon and the Bondi radius is much smaller in this artificial problem, it can be handled by standard GRMHD techniques, e.g., A. Lalakos et al. (2022) presented a simulation with $R_B \approx 10^3\, r_g$. In our tests, we first run the simulation on a single zone (we refer to this as the “one-zone” run hereafter) that covers the entire volume from inside the event horizon to the outermost radius $\approx 10^4\, r_g$. We treat the result from this run as the “true solution.” We then run the same problem with our multizone method using four annuli (“four-zone”). By comparing the results from the one-zone and four-zone simulations, we assess how well the multizone method is able to simulate MHD accretion.

The initial conditions for these test runs are as follows. The density is set to $\rho_{\rm init}(r) \propto (r + R_B)/r$, the temperature is set to the Bondi analytic HD solution, and the velocity is set to zero. Here we only use this single set of initial conditions, as it was demonstrated in Appendix C in H. Cho et al. (2023) that the final steady-state results do not depend on the initial conditions adopted. The magnetic field is initialized with a purely azimuthal vector potential

$$A_\varphi(r, \theta) = \frac{b_z}{2}(r + R_B)\sin\theta, \tag{7}$$

such that the plasma-$\beta \equiv 2\rho T/b^2$ is constant across radii for the initial $\rho$ and $T$ (see Appendix D).[8] The resulting magnetic field geometry is vertical outside $R_B$ and slightly radial inside $R_B$. The $b_z$ parameter determines the strength of the magnetic field.

The same numerical floors are used as in H. Cho et al. (2023) for the MHD runs. The floors are applied in the Eulerian (or normal) observer’s frame, and ensure that the density $\rho > 10^{-6}\, r^{-3/2}$, internal energy $u > 10^{-8}\, r^{-5/2}$, temperature $u/\rho < 100$, and magnetization $b^2/\rho < 100$. We reduce the gas velocity whenever the Lorentz factor measured by the Eulerian observer is larger than $\gamma_{\rm max} = 10$.

The boundary conditions at the innermost and outermost radii, $r_{\rm in}$ and $r_{\rm out}$, are exactly the same for the one-zone and four-zone simulations, namely, free inflow toward $r = 0$ at $r_{\rm in}$ and Dirichlet boundary conditions at $r_{\rm out}$. The one-zone run has no other radial boundaries, whereas the four-zone simulation has interior boundaries between annuli whose boundary conditions follow the description in Section 2.2.

The runtime for each annulus in the four-zone simulations is chosen to be $t_{{\rm MHDcap},i}$ (Equation (6)), for a better comparison with the one-zone simulation. Since the one-zone simulation converges first at the smallest radii and slowly converges out to larger radii, we can barely achieve convergence at the Bondi radius even for this artificial problem with a small $R_B$. By choosing the capped runtime $t_{{\rm MHDcap},i}$, we limit the time spent by the four-zone run at radii $r > R_B$, thereby enabling a closer comparison with the one-zone simulation.

For the following analyses, we define the time-averaged, density-weighted, shell average of a quantity $X$ at a given

[8] There is a typo in the formula for $A_\varphi$ given in H. Cho et al. (2023). Equation (7) here is the correct formula.

radius as:

$$\langle X\rangle(r) = \overline{\left(\frac{\iint X\rho\sqrt{-g}\ d\theta\ d\varphi}{\iint \rho\sqrt{-g}\ d\theta\ d\varphi}\right)}, \quad (8)$$

where the bar represents a time average; $\langle\rho\rangle$ is an exception where it is not additionally density-weighted. The net (inflow) accretion rate is calculated as:

$$\dot{M}(r) \equiv -\iint \rho u^r\sqrt{-g}\ d\theta\ d\varphi, \quad (9)$$

and the feedback efficiency is calculated as (see H. Cho et al. 2023)

$$\eta(r) = \overline{(\dot{M} - \dot{E})}\ /\ \overline{\dot{M}}_{10}, \quad (10)$$

where $\dot{E}(r) \equiv \iint T^r_t\sqrt{-g}\ d\theta\ d\varphi$ is the energy inflow rate at radius $r$ and $\overline{\dot{M}}_{10}$ is the time-averaged accretion rate at $10\,r_g$. When energy is transported outward (inward), the efficiency $\eta$ is positive (negative).

When plotting the radial profiles, the time- and shell-averaged profiles for each zone are stitched together using only the central half of each annulus, i.e., from 25% to 75% of the $\log r$ range, and the remaining regions near the two boundaries are excised. The time averaging is performed over the second half of each visit because the first half of the visit might be affected by the relaxation from the previous active zone's boundary. Both of these choices are made in order to separate boundary artifacts from physical effects. When time-averaged results are presented, they are calculated from the last $1/5$ V-cycles when the solution has reached a quasi-steady state. Finally, when averaging over $\theta$, the last layer of cells closest to each pole is ignored to remove any potential contamination from the reflecting polar boundary conditions. This does not apply to conserved quantities such as the accretion rate $\dot{M}$, $\dot{E}$, or $\eta$ which are summed over all $\theta$.

### 3.1. Weakly Magnetized Bondi Accretion

We first study spherical Bondi accretion initialized with a weak magnetic field: initial plasma-$\beta \approx 10^8$. Since the magnetic field is too weak to affect the dynamics of the gas, the accretion follows basically the H. Bondi solution, with the frozen-in magnetic field being simply advected with the flow. The accretion is axisymmetric, so we run this simulation in two dimensions ($r$, $\theta$) in modified Kerr–Schild (MKS) coordinates (see Appendix C). For the multizone simulation, the resolution is $64^2$ per annulus with a total of $n = 4$ annuli.[9] The corresponding one-zone simulation has the same effective resolution of $160 \times 64$ with $r_{\rm in} = r_g$ and $r_{\rm out} = 8^5\ r_g \approx 3.3 \times 10^4\ r_g$.

The radial profiles of various quantities obtained from the one-zone and four-zone simulations are shown in Figure 2 with solid red lines and dashed orange lines, respectively. The two simulations agree very well with each other and with the analytic Bondi HD solution for the HD quantities, $\dot{M}$, $\rho$, $T$, and $u^r$, as shown in Figures 2(a)–(d). This is expected because as demonstrated previously in H. Cho et al. (2023) the multizone method works well for the Bondi HD problem.

[9] Since the runtimes of the two outermost annuli are both limited by the freefall time at the Bondi radius, we combine these into a larger annulus with $96 \times 64$ resolution. This is also done for the strongly magnetized case, where the largest annulus is run with $96 \times 64^2$ resolution.

Figure 2(e) shows the behavior of the plasma-$\beta$ and is a more significant test because it now involves the magnetic field. The close agreement of the profiles obtained from the one-zone and four-zone simulations indicates that the multizone method handles field advection well despite the presence of internal boundaries between annuli. Note that the plasma-$\beta$ decreases rapidly toward the black hole. Since the magnetic field is advected radially inward, the field strength should scale as $\propto r^{-2}$. Combining this with the Bondi density $\rho \propto r^{-3/2}$ and temperature $T \propto r^{-1}$ (for $r < R_B$), $\beta$ is expected to scale as $\beta \propto r^{3/2}$ (see V. F. Shvartsman 1971). The slope of $r^{3/2}$ is shown with the green line for comparison. The actual slope in the numerical solutions is slightly steeper than $r^{3/2}$, probably because there is only a small range of radii with self-similar behavior (because of the choice of an artificially small Bondi radius for this specific test case).

Note that, even though we used an extremely large initial $\beta = 10^8$ outside the Bondi radius, $\beta$ is $\sim$10 near the BH, and the magnetic field is close to becoming dynamically important. Realistic $\beta$ values for the gas outside the Bondi radius in astrophysical problems are more likely in the range $1-10$, and as we show, they behave very differently. The weak-field case described in this section is thus highly artificial and is of interest only for testing how the algorithm handles field advection.

Figure 2(f) shows the energy outflow efficiency $\eta$ as a function of the radius. For both the one-zone and four-zone simulations, the efficiency is constant as a function of the radius and has a negative value, $\eta \approx -4 \times 10^{-3}$. This is consistent with the estimate for spherically symmetric HD accretion (Equation (2) of H. Cho et al. 2023),

$$\eta_{\rm HD} \rightarrow -\gamma_{\rm ad}T_\infty/(\gamma_{\rm ad} - 1) = -1.5 r_g/R_B \approx -4 \times 10^{-3}. \quad (11)$$

The negative efficiency implies that there is no net energy outflow (no feedback).

In summary, the weak magnetic field simulations converge to the same solution independent of the number of zones used and match the relativistic Bondi analytic solution. The magnetic field flows in with the gas and the Bondi accretion rate $\dot{M}_B$ is recovered. In this case, there is no feedback.

### 3.2. Strongly Magnetized Bondi Accretion

We now repeat the same test using a much stronger initial magnetic field: $\beta_{\rm init}(r) \sim 1$. It is known from previous studies (e.g., I. V. Igumenshchev & R. Narayan 2002; I. V. Igumenshchev et al. 2003; R. Narayan et al. 2003; M. C. Begelman et al. 2022) that the nonaxisymmetric Rayleigh–Taylor instability (or interchange instability) is activated in the presence of strong magnetic fields in MADs. Therefore, we run these simulations in three dimensions to fully capture the instability, and use our new wide-pole Kerr–Schild (WKS) coordinate system (Appendix C). The four-zone simulation has a resolution of $64^3$ per annulus and the one-zone has the same effective resolution of $160 \times 64^2$. In order to break the symmetry of the initial axisymmetric equilibrium, initial random perturbations are applied to the internal energy at all radii, with a maximum amplitude of 10%.

Converged radial profiles of various quantities for this strongly magnetized Bondi accretion test are shown in Figure 2, with the one-zone and four-zone runs shown as solid black lines and dashed cyan lines, respectively. The two simulations are in very good agreement with each other across all compared quantities. Therefore, the multizone method passes this challenging test successfully.

Looking at the results in detail, when strong magnetic fields are present, (i) the accretion rate is reduced relative to the H. Bondi rate, (ii) the density $\rho$ deviates from the $r^{-3/2}$ prediction of the Bondi analytical solution and scales instead as $\rho \propto r^{-1}$, (iii) there is a temperature bump at around $r \approx R_B$, and (iv) the outflow efficiency $\eta$ is positive, implying that there is a net energy outflow. All these results are very similar to those presented in H. Cho et al. (2023) for the highly magnetized Bondi accretion case (except that those simulations used a realistic Bondi radius of $R_B \gtrsim 10^5 r_g$ whereas the present simulations have $R_B \approx 400 r_g$). We also note that the density scaling of $\rho \propto r^{-1}$ is consistent with previous simulation studies (S. M. Ressler et al. 2018, 2020; K. Chatterjee & R. Narayan 2022; M. Guo et al. 2023, 2024), the analytical model of W. Xu (2023), and observational constraints from M87* and Sgr A* (H. R. Russell et al. 2015; K. Chatterjee & R. Narayan 2022).

The demonstration of a positive outward flux of energy (feedback) in strongly magnetized Bondi accretion was one of the highlights of H. Cho et al. (2023). The confirmation of this effect here, for both the "true" one-zone solution and the four-zone simulation, is significant. The feedback efficiency, $\eta \sim 1\%-2\%$, is also similar. As shown in H. Cho et al. (2023), the outward energy flux is carried by a form of magnetized convection. The bump in the temperature at $r \approx R_B$ (Figure 2(c)) is the result of the outflowing energy being deposited near the Bondi radius.

In H. Cho et al. (2023), the accretion rate was suppressed relative to the Bondi rate by 2 orders of magnitude, whereas here it is suppressed by only 1 order of magnitude. Since the only difference between the two simulations is the choice of Bondi radius, $R_B/r_g \approx 2 \times 10^5$ versus 400, we see that the accretion rate (not surprisingly) depends on $R_B$. We explore this further in Section 6.

## 4. Hydrodynamic Accretion with Boundary Conditions from a Galaxy-scale Simulation

The classical hydrodynamic Bondi accretion problem discussed in H. Cho et al. (2023) assumes a homogeneous, constant density/temperature external medium, which extends out to infinity with no self-gravity. This is highly idealized. A more realistic scenario is to consider spherical accretion from an external medium that is in hydrostatic equilibrium in the gravitational potential of a galactic nucleus. Here, we simulate spherically symmetric three-dimensional (3D) GRHD accretion in such a scenario.

### 4.1. GIZMO Galaxy HD Simulation Setup

In order to provide realistic outer boundary conditions for the GRHD simulation, we simulate an isolated galaxy with an M87-like dark matter halo using the GIZMO[10] code (P. F. Hopkins 2015). For this run, we use GIZMO in its meshless finite mass mode. It is a Lagrangian mesh-free Godunov method, capturing both the advantages of grid-based and smoothed-particle hydrodynamics methods. Numerical details and extensive tests are described in a series of previously published methods papers for, e.g., the HD and self-gravity (P. F. Hopkins 2015), MHD (P. F. Hopkins 2016; P. F. Hopkins & M. J. Raives 2016), anisotropic Spitzer–Braginskii conduction and viscosity (P. F. Hopkins 2017; K.-Y. Su et al. 2017).

In this initial application, we run the galaxy-scale simulation purely adiabatically for 100 Myr, without any cooling. Baryonic processes like star formation and stellar feedback are also not included, as our immediate goal is to extract the appropriate boundary conditions for use at the outer radii of the GRMHD simulations.

Initial conditions for the galaxy-scale simulation resembling an M87-like host galaxy are adopted, with the dark matter (DM) halo, stellar bulge, BH, and gas halo initialized following procedures described in V. Springel & S. D. M. White (1999), V. Springel (2000), and K.-Y. Su et al. (2019, 2020, 2021, 2024). We start with a spherical, isotropic DM halo with a Navarro–Frenk–White profile (J. F. Navarro et al. 1996) of a halo mass of $1.97 \times 10^{14} M_\odot$ and a scale radius of 448 kpc (L. J. Oldham & M. W. Auger 2016). The stellar bulge follows an L. Hernquist (1990) profile with a mass of $6.87 \times 10^{11} M_\odot$ and a scale radius of 3.97 kpc, roughly fitting the observation of J. C. Forte et al. (2012). The BH has a mass of $M_\bullet = 6.5 \times 10^9 M_\odot$. The gas in the halo is in hydrostatic equilibrium and follows a $\beta$-profile with a mass of $4.7 \times 10^{13} M_\odot$, $\beta = 0.33$, and a scale radius of 0.93 kpc (E. Churazov et al. 2008). The gas in the halo has a rotation of $0.1\Omega_K$, where $\Omega_K$ is the Keplerian angular velocity and gets most of its support from thermal pressure, as expected in such massive halos. The mass resolutions of DM, stars, and gas are, respectively, $8 \times 10^6 M_\odot$, $3 \times 10^4 M_\odot$, and $3 \times 10^4 M_\odot$.[11]

### 4.2. KHARMA GRHD Simulation Setup

After evolving the GIZMO simulation for 100 Myr, the shell-averaged $\rho(r)$, $T(r)$ profiles are used to initialize the outermost zone 7 ($r \approx 10^6-10^8\, r_g$). This is to retain spherical symmetry that will permit direct comparison with our prior results on the H. Bondi accretion problem presented in H. Cho et al. (2023). In a similar spirit, we initialize the gas with zero velocity even though GIZMO gives a nonzero shell-averaged velocity. This will soon be generalized to nonspherically symmetric initial conditions in the MHD run discussed in Section 5. The temperature at a large radius $T_\infty$ from GIZMO corresponds to a realistic Bondi radius of $R_B \approx 2 \times 10^5\, r_g$. The runtime for zone $i$ is $t_{\mathrm{HD},i}$ (Equation (4)), and no numerical floors are used. The coordinate system is exponential Kerr–Schild (eKS; see Appendix C).

The gravitational effect of the galaxy is included by modifying the Kerr–Schild metric to the following (see Equation (B4), Appendix B),

$$ds^2 = -\left(1 - \frac{2}{r} + 2\Phi_g\right) dt^2 + 4\left(\frac{1}{r} - \Phi_g\right) dt\, dr + \left(1 + \frac{2}{r} - 2\Phi_g\right) dr^2 + r^2\, d\Omega^2, \qquad (12)$$

[10] A public version of GIZMO is available at http://www.tapir.caltech.edu/~phopkins/Site/GIZMO.html.

[11] We use a hierarchical super-Lagrangian refinement scheme (K.-Y. Su et al. 2019, 2021, and K.-Y. Su et al. 2024) to reach $\sim 3 \times 10^4\, M_\odot$ mass resolution in the core region and around the $z$-axis where we expect a jet will be launched for future work. The mass resolution decreases as a function of both the radius ($r_{3\mathrm{d}}$) and distance from the $z$-axis ($r_{2\mathrm{d}}$), proportional to $r_{3\mathrm{d}}$ and $2^{r_{2\mathrm{d}}/10\,\mathrm{kpc}}$, whichever is smaller in mass, to $\sim 2 \times 10^6 M_\odot$. The highest resolution is manifested in regions where either $r_{3\mathrm{d}}$ or $r_{2\mathrm{d}}$ is smaller than 10 kpc.

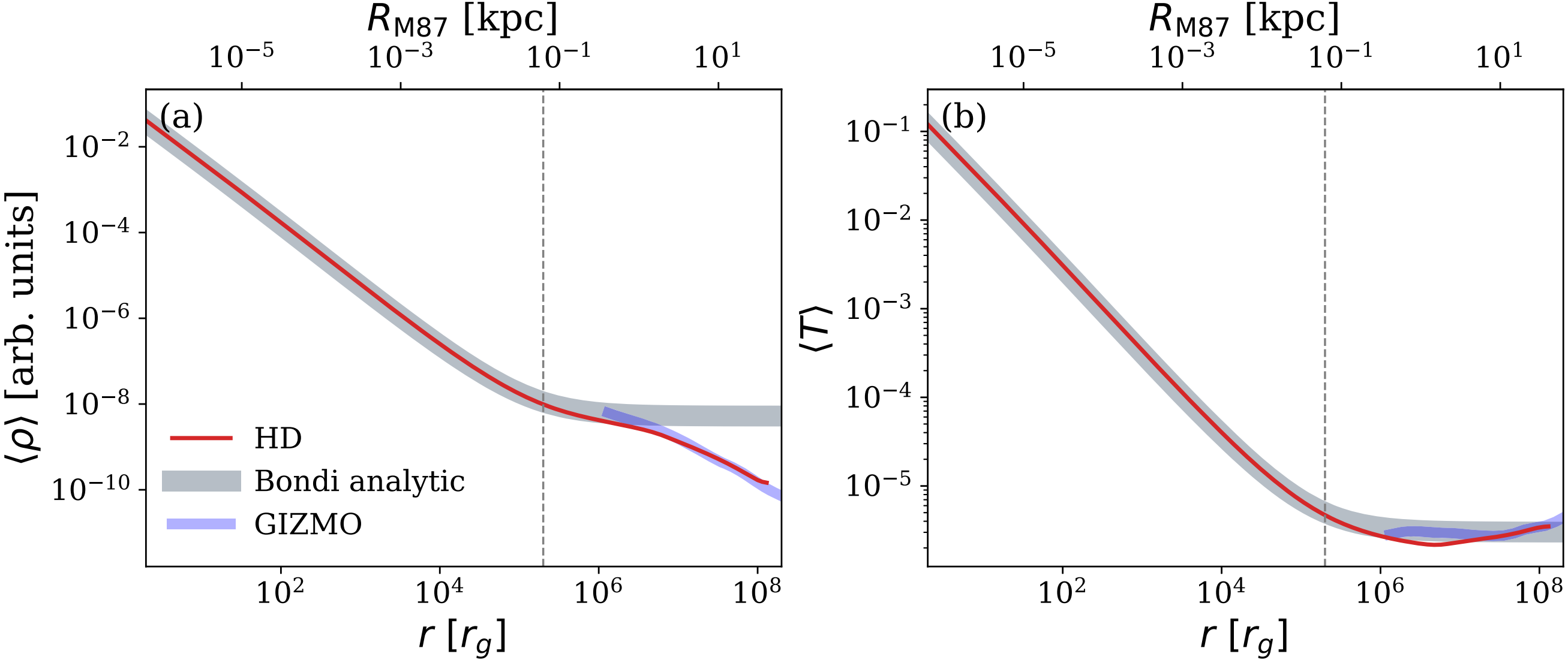

**Figure 3.** Time-averaged radial profiles of (a) density $\rho$, and (b) temperature $T$, for a GRHD spherically symmetric accretion simulation where large radii are initialized from a realistic galaxy simulation using the code GIZMO. Radii are expressed in gravitational radius $r_g$ units (lower $x$-axis) and in physical units scaled to M87 $R_{\rm M87}$ (upper $x$-axis). The GRHD simulation includes the external gravity of the galaxy, and the Bondi radius is $R_B \approx 2 \times 10^5 r_g$ (indicated by the gray vertical line). The initial data from GIZMO are shown for $r \geqslant 10^6 r_g$ by the thick blue lines. The Bondi analytical solution is shown by the thick gray lines. Results from our eight-zone GRHD simulation, shown by the red lines, match the Bondi solution inside the Bondi radius and closely follow the GIZMO profiles at larger radii. A similar run which is initialized with a strong magnetic field gives very different results, as shown in Figure 5.

where $\Phi_g(r)$ is the gravitational potential of the galaxy, and $d\Omega^2 = d\theta^2 + \sin^2\theta\, d\varphi^2$. The external gravitational potential $\Phi_g(r)$ is calculated by measuring the enclosed mass at radii $\sim$0.3–30 kpc (or $10^6$–$10^8\ r_g$) from the GIZMO run. While the enclosed mass profile includes the contributions of all the constituents—BH, DM, stars, and gas—it is the stars that dominate the gravitational potential on these scales. The total enclosed mass is then fitted with a power law and integrated to obtain an approximation for the external gravitational potential:

$$\Phi_g(r) \approx 10^{-8}(r^{0.36} - (2\ r_g)^{0.36}). \tag{13}$$

We have set the zero-point of the galaxy potential at the horizon radius, $r_H = 2\ r_g$. As explained in Appendix B, this has the convenient property of not changing the location of the horizon even after the galaxy potential is explicitly included in the metric.

### 4.3. Results

In Figure 3, we plot the time- and shell-averaged $\rho(r)$ and $T(r)$ obtained from the KHARMA simulation in red solid lines. The initial $\rho(r)$ and $T(r)$ from the GIZMO run are shown as thick blue solid lines. The Bondi analytical solution normalized to match the GIZMO computed values at $r = 10^6\ r_g$ is shown as thick gray lines. Unsurprisingly, we find that within the region of influence of the BH, $r < R_B$, where the BH's gravity dominates, the inclusion of external gravity has no impact, and the analytic Bondi solution is recovered as expected. However, on scales beyond $R_B$, the solution undergoes a transition. In the classical H. Bondi problem, the density and temperature asymptote to constant values at large $r$. Here instead, $\rho$ matches smoothly onto the galaxy density profile obtained from GIZMO, which is the correct solution for the assumed galaxy potential. The density $\rho$ decreases at large $r$ and this is a direct result of incorporating the external gravitational potential of the galaxy.

The accretion rate obtained in this simulation is $\dot{M} = \dot{M}_B$, and the efficiency is $\eta \sim -1.5 r_g/R_B \sim 8 \times 10^{-6}$. Both are consistent with the Bondi HD results reported in H. Cho et al. (2023).

## 5. Magnetohydrodynamic Accretion Initialized from a Galaxy Simulation

To get a step closer to simulating accretion from a realistic galaxy, we now include strong magnetic fields and nonspherically symmetric initial conditions for the gas, again using a GIZMO simulation with external gravity and $R_B \approx 2 \times 10^5\ r_g$.

### 5.1. GIZMO MHD Simulation Setup

The initial condition of the GIZMO MHD run is mostly identical to the HD case described in Section 4.1. The initial magnetic fields are dominated by a poloidal component with $\beta \approx 10$ for $r < 10$ kpc and smoothly decay to a toroidally dominated magnetic field of $\beta \approx 1000$ at distances $r > 10$ kpc. The detailed form of the magnetic field is described in Appendix D. Once again, as done for the HD case, we ran the GIZMO simulation for 100 Myr.

To provide 3D initial conditions for the subsequent KHARMA run, we deposit the density field onto the combined global KHARMA grid covering zones 0–7 with a cubic spline kernel. According to the mass contribution of each resolution element from the GIZMO snapshot to each KHARMA grid cell, we mass-weight average the temperature, velocity, and magnetic fields and deposit on the KHARMA grid.

### 5.2. KHARMA GRMHD Simulation Setup

Unlike the HD simulation described in Section 4.2 where shell-averaged GIZMO data were used, we now use the full 3D gas and magnetic field data from GIZMO as functions of $(r, \theta, \varphi)$ to initialize the KHARMA GRMHD simulation. The gas also has a small net angular velocity $\Omega \sim 0.1\Omega_K$ at all radii.

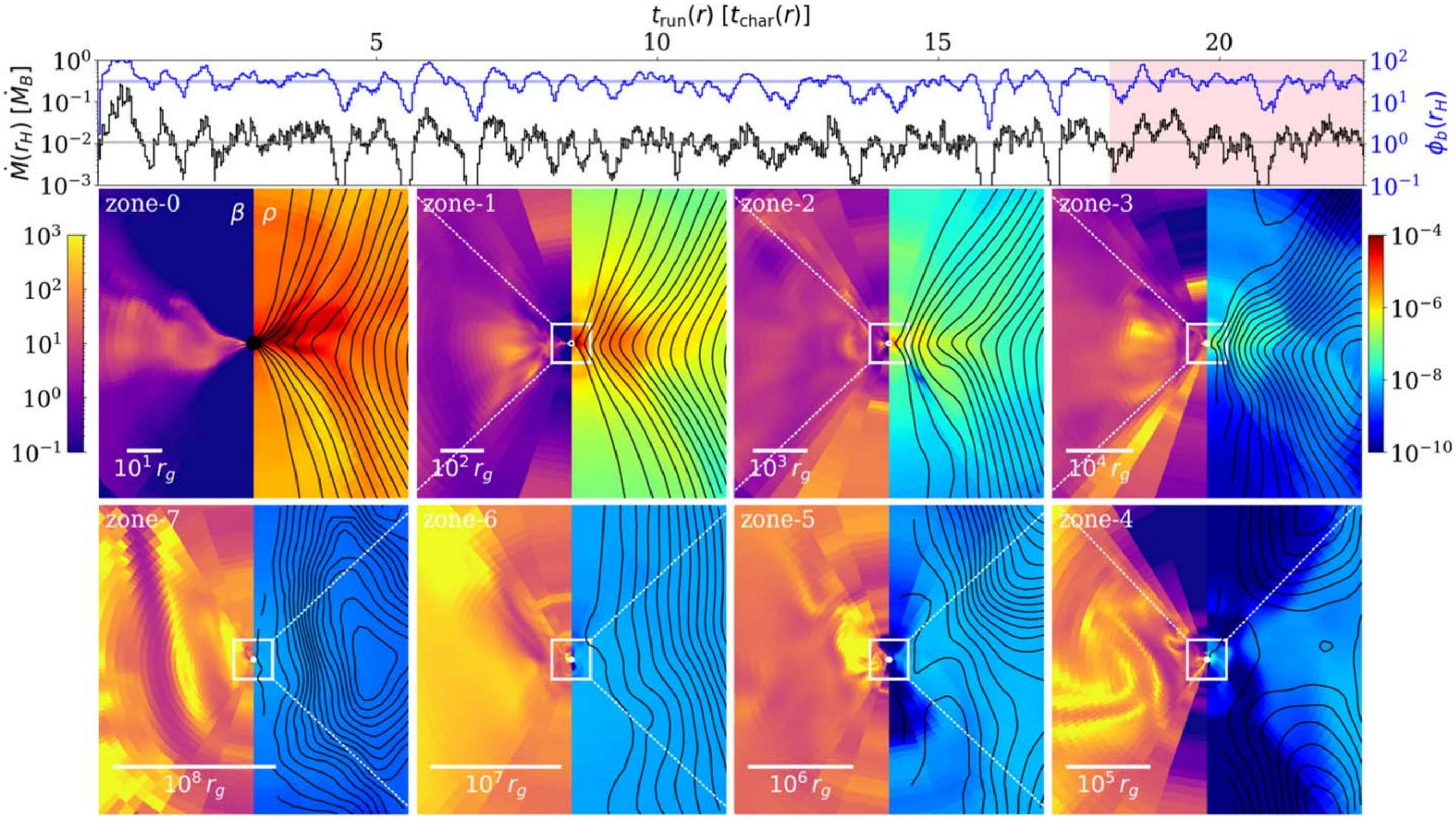

**Figure 4.** Top: accretion rate (black) and magnetic flux parameter (blue) at the horizon $r_H$ as a function of time for a multizone GRMHD simulation initialized with GIZMO galaxy simulation data. Time is in units of the characteristic time $t_{\rm char}(r)$ spent at each zone. The mean accretion rate, $\overline{\dot{M}} \sim 10^{-2}\dot{M}_B$, and the mean magnetic flux at the horizon, $\overline{\phi_b} \sim 30$, are plotted in gray and blue horizontal lines, respectively. We average values over the last $1/5$ of the simulations (pink background) when the simulation has reached a steady state. Bottom: depiction of the accretion flow across multiple radial scales from a single late-time snapshot of the multizone simulation. In each panel, the right half shows the distribution of the density $\rho$ (color scale on the right) and the left half shows the distribution of plasma-$\beta$ (color scale on the left). The black lines show the magnetic field. Note that turbulent fluctuations extend over 8 orders of magnitude in radius.

The deposited magnetic fields adopted from GIZMO for the KHARMA run have negligible divergence $(\sqrt{-g}\,B^i)_{,i}/\sqrt{-g} < 10^{-7}$ so we do not additionally perform divergence cleaning. By not cleaning, the magnetic field configuration from GIZMO is preserved without introducing significant distortions in the field lines.

For the simulations described in this Section (and for most other simulations in the rest of the paper), we use the runtime prescription $t_{{\rm MHD},i}$ (Equation (5)).[12] We use the WKS coordinate system. The numerical floors are the same as in the small-scale runs outlined in Section 3, except that for the density and internal energy density we use $\rho > 10^{-8} r^{-3/2}$, $u > 10^{-14} r^{-5/2}$. The metric is given in Equation (12), which includes the gravitational potential of the galaxy.

### 5.3. Results

The accretion rate $\dot{M}$ and the dimensionless magnetic flux parameter $\phi_b(r) \equiv \sqrt{\pi/\overline{\dot{M}}_{10}} \iint |B^r| \sqrt{-g}\; d\theta\; d\varphi$ at the horizon obtained from the KHARMA multizone GRMHD simulation are shown as functions of time in the top panel of Figure 4. As in the GRMH. Bondi run in H. Cho et al. (2023), we find $\dot{M} \sim 10^{-2}\dot{M}_B$ and $\phi_b \sim 30$.[13] In making these estimates, we redefine the analytic Bondi accretion rate $\dot{M}_B$ to be

$$\dot{M}_B \rightarrow \dot{M}_B \frac{\langle\rho\rangle(r = R_B)}{\rho_\infty}, \tag{14}$$

where $\rho_\infty$ is the asymptotic density at infinity of the analytic solution. This adjustment is made to compensate for the overall density shift at the Bondi radius (note the difference of $\rho(R_B)$ between the analytic solution and the numerical result in Figure 5(a)). This can be thought of as replacing the boundary condition $\rho_\infty$ with the actual numerical value of $\langle\rho\rangle(R_B)$ in the Newtonian $\dot{M}_B$ formula (Equation (A1)).

The bottom two rows of Figure 4 show a snapshot of the simulation. Compared to the GRMH. Bondi accretion solution shown in H. Cho et al. (2023), which had nearly homogeneous distributions of density and magnetic field lines in zones 5–7, in the example shown here the distributions are highly nonuniform even at the largest radii $r \gg R_B$. This is in part because we used nonuniform initial conditions from GIZMO, but more importantly, because the present simulation has been run for a longer time outside $R_B$ as a result of using the $t_{\rm MHD}$ runtime prescription instead of the $t_{\rm MHDcap}$ prescription used in H. Cho et al. (2023).

The time- and shell-averaged radial profiles of several quantities are shown in Figures 5(a)–(d).[14] For $r < R_B$, these results are consistent with the results shown in H. Cho et al.

[12] In Section 5.4, we show that different choices of the runtime have negligible effect on the results for $r < R_B$.

[13] H. Cho et al. (2023) reported $\dot{M} \approx 0.005\dot{M}_B$ for their GRMHD simulation. However, when we adjust the estimate of $\dot{M}_B$ as in Equation (14), the H. Cho et al. (2023) run also gives $\dot{M} \approx 0.01\dot{M}_B$.

[14] The profiles have been smoothed radially to reduce noise fluctuations.

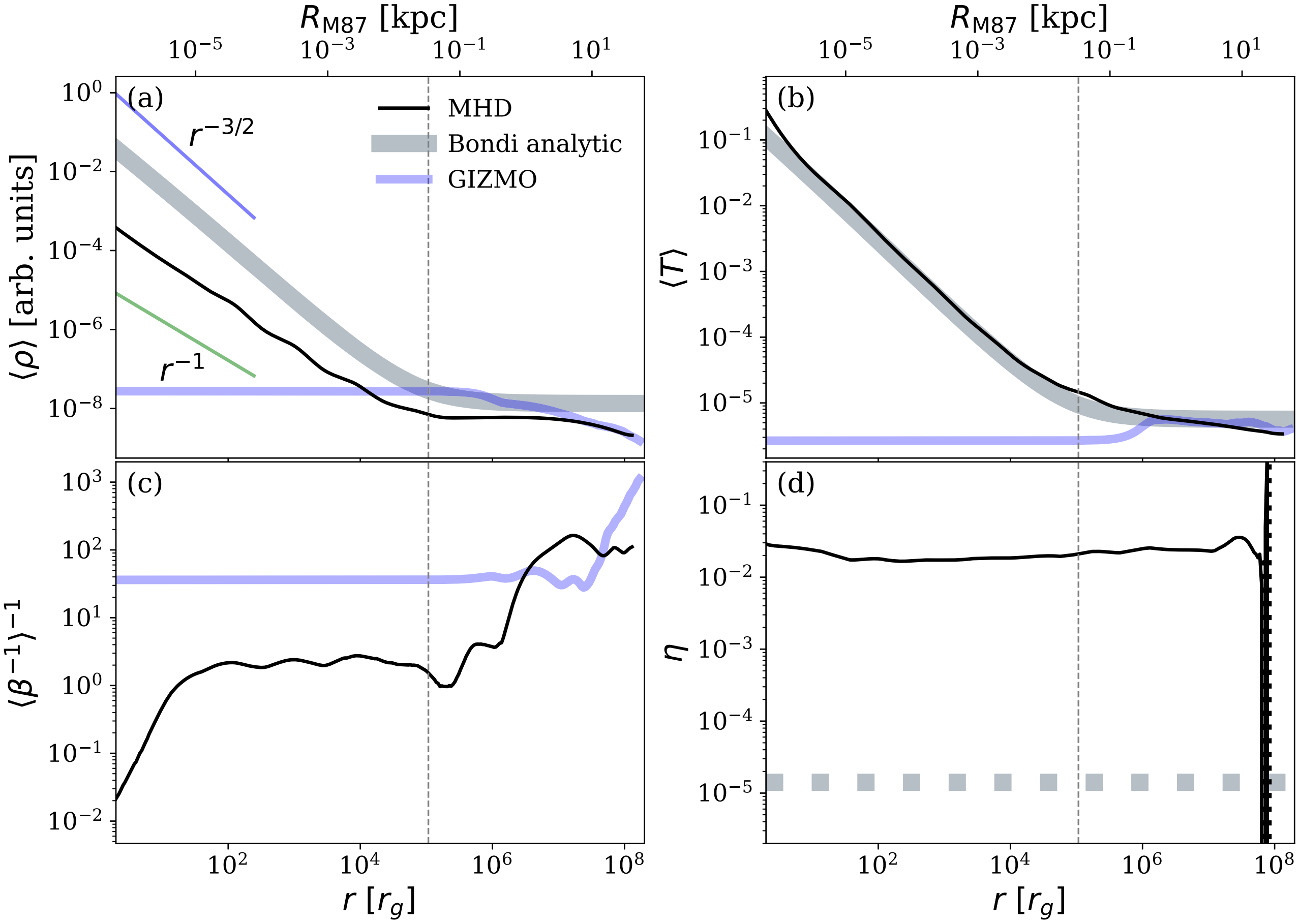

**Figure 5.** Time-averaged radial profiles of (a) density $\rho$ (b) temperature $T$ (c) plasma-$\beta$, and (d) efficiency $\eta$, of a multizone GRMHD simulation initialized from GIZMO data, including the external gravity of the galaxy. The initial data from GIZMO are shown as thick blue lines and the Bondi analytical solution, normalized to the initial GIZMO density at $r \approx 10^6\, r_g$, is shown in thick gray lines. Because of feedback from strong magnetic fields, the density in the radius range $R_B - 100R_B$ in the multizone GRMHD simulation (black solid lines) deviates noticeably from the initial density profile. Also, the feedback efficiency of a few percent ($\eta \sim 2\%$) is constant over 7 orders of magnitude in spatial scales. Both effects indicate that the feedback from the accreting BH is impacting the gas dynamics on large galactic scales. The temperature closely follows the Bondi solution and the plasma-$\beta$ profile saturates to a value of order a few between $r = 30-10^5\, r_g$ before decreasing toward the BH. Overall, the results for $r < R_B$ are similar to the simple GRMH. Bondi problem described in H. Cho et al. (2023).

(2023) for their strongly magnetized Bondi accretion simulation. In detail, the density has converged to a slope of $r^{-1}$, the temperature closely follows the Bondi analytic solution, the plasma-$\beta$ has saturated at a value of order unity from $r = 10^{1.5}-10^5\, r_g$ (also consistent with S. M. Ressler et al. 2023), and decreases closer to the BH, and the feedback efficiency $\eta$ is of order a few percent. Even though the initial conditions are quite different in the two works (weaker and more inhomogeneous initial magnetic field, 3D fluctuations in gas initial density, temperature, velocity, external gravity of the galaxy, etc., in the simulation described here), we obtain more or less the same results. This illustrates that, when the BH is provided with enough magnetic flux in the external medium, the steady state of the magnetically saturated accretion flow for a nonspinning black hole is insensitive to the initial conditions. This statement applies to the region of the solution interior to the Bondi radius, $r < R_B$.

At larger radii, $r > R_B$, the final density profile (black solid line) in the present simulation has evolved from its initial state (shown by the blue band), contrary to the corresponding HD simulation (Figure 3(a)) or the GRMHD simulation in H. Cho et al. (2023). This indicates that the current multizone GRMHD simulation exhibits strong feedback effects on scales well beyond $R_B$, moving the system away from its original hydrostatic equilibrium. Evidence for this can also be seen in Figure 5(d), where the constant positive feedback efficiency $\eta$ extends all the way to $\sim 10^{7.5}\, r_g$ (which corresponds to about 10 kpc in M87).

Surprisingly, the presence of nonzero angular velocity, $\Omega \sim 0.1\Omega_K$, in the GIZMO-provided initial conditions has barely any effect on our GRMHD results. We had anticipated that angular momentum conservation would cause the accreting gas to spin up to $\Omega/\Omega_K \sim 0.5-1$ at smaller radii and that the accretion flow would maintain this level of rotation down to the BH, just as in standard GRMHD simulations which are initialized with rotating gas in a torus (e.g., R. Narayan et al. 2012). We had further expected the efficiency $\eta$ to increase modestly above the $\sim 2\%$ that we obtain for nonrotating initial conditions. Instead, what we find is that the angular velocity never becomes important at any radius, and the efficiency is hardly modified. When tracking the angular velocity evolution, we noticed that $\Omega$ flips its sign frequently at all radii even though the gas is globally initialized with $\Omega$ of a single sign. As a result of this flip-flopping, the time-averaged angular velocity profile resembles that of the nonrotating MHD simulation. Note that, while R. Narayan et al. (2012) report an ordered rotation for a nonspinning BH, their $\Omega/\Omega_K$ continuously decreases over time in the MAD state. It is possible that $\Omega$ will eventually go

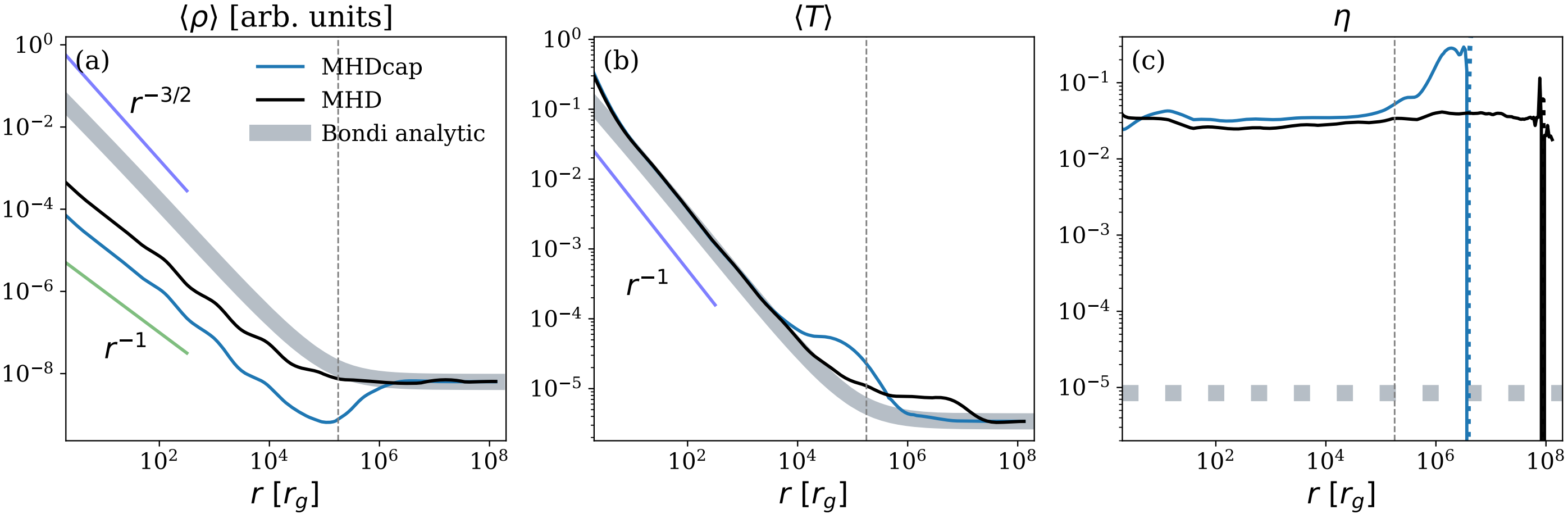

**Figure 6.** Comparison of the results from multizone GRMHD simulations using two different prescriptions for the runtime per annulus: $t_{\rm MHD}$ (black lines) in which the runtime is proportional to the local characteristic timescale of the annulus, and $t_{\rm MHDcap}$ (blue lines), which is similar but capped at the Bondi timescale $t_B$. The panels show radial profiles of (a) density $\rho$ (b) temperature $T$, and (c) feedback efficiency $\eta$. Because $t_{\rm MHDcap}$ limits the runtime at radii larger than $R_B$, the outflowing energy from feedback piles up at the Bondi radius, creating a high-temperature low-density region around $R_B$. On the other hand, the simulation with $t_{\rm MHD}$ allows enough time for the density dip to be smoothed out and for the temperature bump from feedback to propagate farther out to $10^7\ r_g \sim 100\ R_B$. The positive feedback efficiency $\eta$ in this run is constant over nearly 8 orders of magnitude in radius, all the way from the event horizon to around 100 $R_B$, whereas with the $t_{\rm MHDcap}$ prescription, $\eta$ is constant only up to $\approx R_B$. Regardless, both simulations agree on the most important results, namely, the density scales as $\rho \propto r^{-1}$ inside the Bondi radius, the accretion rate on the BH $\dot{M} \approx 0.01\ \dot{M}_B$, and the feedback efficiency $\eta \approx$ a few percent.

to zero when the simulation is run long enough, consistent with our findings.

A natural suspicion is that these counterintuitive results with rotating initial conditions may be an artifact of our multizone approach, especially the Dirichlet boundary conditions between annuli. We discuss this question further in Section 7.

### 5.4. Effect of Different Runtime Per Zone

We introduced two different prescriptions for the runtime per zone in Section 2.3: $t_{{\rm MHD},i}$ (Equation (5)) and $t_{{\rm MHDcap},i}$ (Equation (6)). All the MHD simulations in H. Cho et al. (2023) and the simulations with small Bondi radius $R_B < 10^5\ r_g$ in the present paper (Sections 3 and 6) use $t_{{\rm MHDcap},i}$. The remaining simulations in this work, with $R_B > 10^5\ r_g$ (e.g., Figure 5), use $t_{{\rm MHD},i}$. The main difference between the two runtime prescriptions is that $t_{{\rm MHDcap},i}$ limits the amount of runtime spent at radii $>R_B$, while $t_{{\rm MHD},i}$ allows all zones to run proportional to their own characteristic timescales $t_{\rm char}$. In this section, we study how the choice of runtime affects the results.

For this test, we return to the basic GRMH. Bondi accretion problem presented in H. Cho et al. (2023), namely, homogeneous constant density/temperature external medium, and no external gravity. The initial density is $\rho_{\rm init}(r) \propto (r + R_B)/r$ and the initial plasma-$\beta$ is of order unity. We use the pure Schwarzschild metric ($\Phi_g = 0$ in Equation (12)) in WKS coordinates. Figure 6 compares the $t$-, $\theta$-, $\varphi$-averaged profiles of density $\langle\rho\rangle(r)$, temperature $\langle T\rangle(r)$, and efficiency $\eta$ ($r$) from two multizone GRMHD simulations of this test problem, one using $t_{{\rm MHD},i}$ (black lines, called `MHD` run hereafter) and the other using $t_{{\rm MHDcap},i}$ (blue lines, called `MHDcap` run hereafter).

We begin by discussing the major differences between the `MHD` and `MHDcap` runs. In Figures 6(a), (b), the `MHDcap` profile exhibits a density suppression and a temperature bump at around $\sim R_B$. In contrast, the density in the `MHD` run is smooth through the Bondi radius and the temperature bump appears farther out at $\sim 100\ R_B$. In the `MHDcap` run, the energy feedback has had little time to propagate to larger radii or to dissipate because this runtime prescription gives little time for larger radii to evolve. The hot gas thus accumulates around $r \sim R_B$ and creates a region of high temperature and low density. Meanwhile, in the `MHD` run, each zone is evolved for a sufficient time relative to its own characteristic timescale $t_{\rm char}$. This lets the dip in the gas density outside the Bondi radius flatten out. Correspondingly, the temperature bump also moves farther out. A similar explanation applies to the feedback efficiency $\eta$ profiles of the two simulations in Figure 6(c). The `MHD` run shows a constant positive $\eta$ profile extending from the horizon to almost $\sim 10^8\ r_g$, while the `MHDcap` run has a constant $\eta$ only out to $R_B$. Note that the computer time needed for the `MHD` run is only slightly longer than that needed for the `MHDcap` run. Thus, with our multizone method, not only can we bridge the region between the BH and the Bondi radius (which was the point of H. Cho et al. 2023), we can push far out into the galaxy if needed.

Despite the differences between the `MHD` and `MHDcap` runs described above, several important results are independent of which runtime prescription is used. Even though the density dip at $R_B$ in the `MHDcap` run introduces an overall offset in the density profiles of the two simulations interior to $R_B$, the density slopes are identical, $\rho \propto r^{-1}$. For estimating the accretion rate on the BH, when we adjust the Bondi rate $\dot{M}_B$ as in Equation (14), using the actual density at $R_B$ rather than the initial density, both runs give the same result for the BH accretion rate: $\dot{M} \approx 0.01\dot{M}_B$. Similarly, the feedback efficiency parameter is almost the same in the two simulations: $\eta$ of a few percent.

## 6. Mass Accretion Rate as a Function of the Bondi Radius

A strongly magnetized system with a small Bondi radius, $R_B \approx 400\ r_g$ (Section 3.2), and a system with a realistic Bondi radius, $R_B \approx 2 \times 10^5\ r_g$ (H. Cho et al. 2023), show very similar steady states, with density scaling as $\rho \propto r^{-1}$ interior to $R_B$, plasma-$\beta$ parameter saturating at order unity across a large range of radii (MAD state), and feedback efficiency converging to $\eta \approx 2\%$. However, the accretion rate $\dot{M}$, scaled to the Bondi rate $\dot{M}_B$, is an order of magnitude less in the case of the larger Bondi radius, $R_B \approx 2 \times 10^5\, r_g$, compared to $R_B \approx 400\, r_g$ (Section 3.2). Since the multizone method enables us to simulate GRMHD accretion for any choice of $R_B$, here we compare a total of seven models with Bondi radii ranging from $R_B \approx 10^2\, r_g - 10^7\, r_g$.

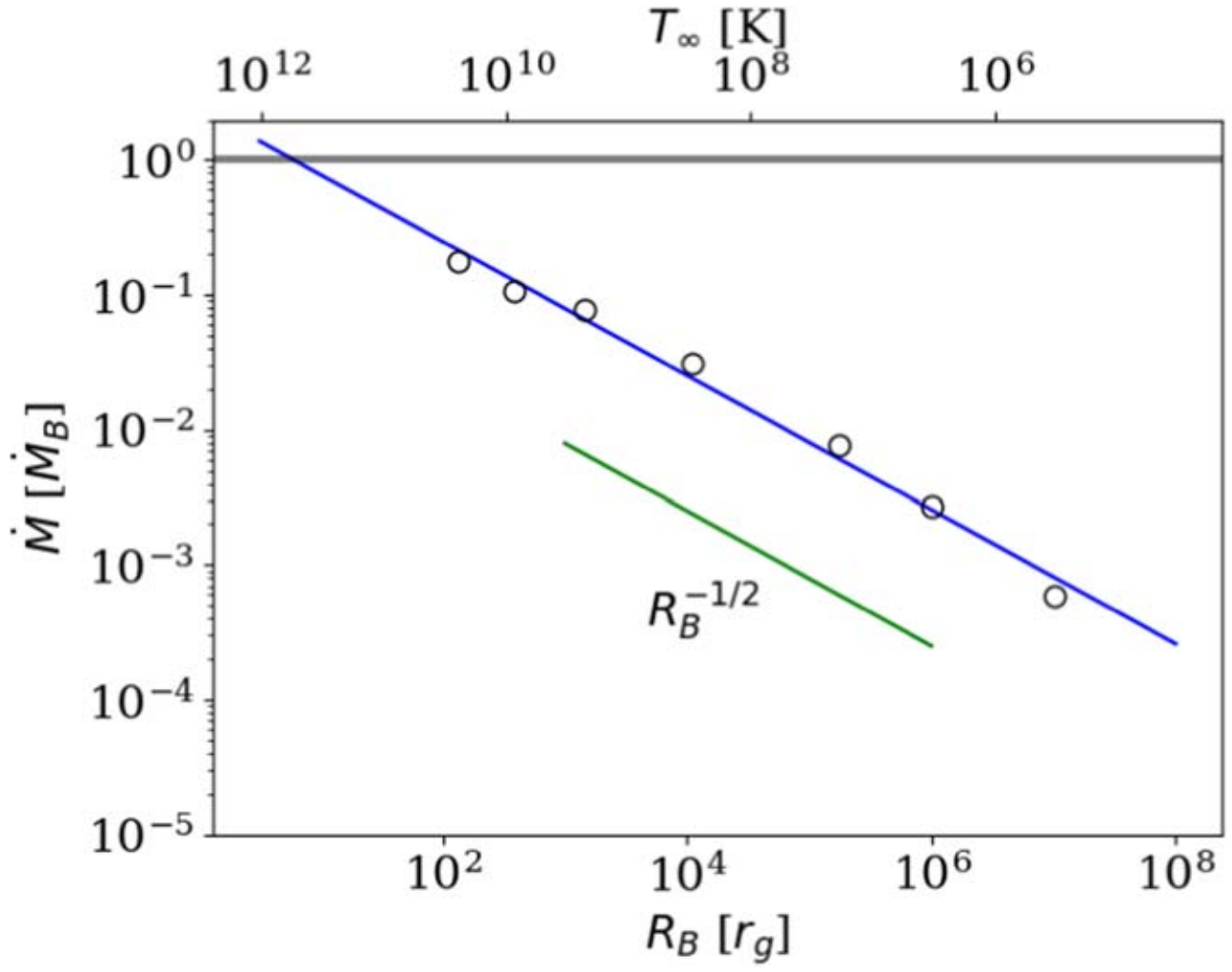

**Figure 7.** Converged BH accretion rate $\dot{M}$ in units of the Bondi accretion rate $\dot{M}_B$ from multizone GRMHD simulations with different values of the Bondi radius $R_B$. The accretion rate is suppressed more for larger values of $R_B$. The naive expectation, $\dot{M} = \dot{M}_B$, is shown as a gray horizontal line, and the fitted power law, Equation (15), is shown as a blue line.

The sonic radii $r_s$ for the seven simulations are $r_s/r_g =$ 10, 16, 30, 80, $\sqrt{10^5}$, 750, 2400, which correspond to Bondi radii, $R_B/r_g \approx 100$, 400, 1500, $10^4$, $2 \times 10^5$, $10^6$, $10^7$, respectively. The number of zones is 4, 4, 4, 5, 8, 9, and 10, respectively, in the seven runs. For the simulations with $R_B < 10^5\, r_g$, we used the `MHDcap` runtime prescription, and for the larger runs, we used the `MHD` prescription. For each model, we measured the steady-state accretion rate $\dot{M}$ on to the BH and the Bondi accretion rate $\dot{M}_B$ corresponding to the final steady-state density profile (see Equation (14)).

In Figure 7, the scaled accretion rates $\dot{M}/\dot{M}_B$ of the simulations are shown as a function of the Bondi radius $R_B$. There is a clear trend of decreasing $\dot{M}/\dot{M}_B$ with increasing $R_B$. The best-fit power law (shown in a blue line) is

$$\frac{\dot{M}}{\dot{M}_B} \approx \left(\frac{R_B}{6\, r_g}\right)^{-0.50}. \tag{15}$$

Since the density interior to the Bondi radius scales as $\rho \propto r^{-1}$ in our strongly magnetized GRMHD simulations, instead of $r^{-3/2}$ as predicted by the analytical Bondi solution, we expect the accretion rate at the BH to be suppressed relative to the Bondi rate by roughly a factor $\sim (r_H/R_B)^{1/2}$, i.e., $\dot{M}/\dot{M}_B \approx (R_B/2r_g)^{-1/2}$. The predicted slope of $-1/2$, shown as a green solid line in Figure 7, matches the simulation results very well. The radius $6\, r_g$ at which Equation (15) predicts $\dot{M}/\dot{M}_B = 1$, might possibly be set by $R_{\rm ISCO}$, the radius of the innermost stable circular orbit ($=6r_g$ for a nonspinning BH), although as we discuss in the next section, rotation does not appear to play much of a role in these solutions. We note that using nested Newtonian MHD simulations, M. Guo et al. (2024) independently predicted that the accretion rate scales with the Bondi radius as $\dot{M}/\dot{M}_B \propto (R_B/r_g)^{-1/2}$.

We note that A. Lalakos et al. (2022) report a larger suppression of the accretion rate, $\dot{M}/\dot{M}_B \approx 0.02$, for their simulation with $R_B = 1000\, r_g$, compared to Equation (15), which predicts $\dot{M}/\dot{M}_B \approx 0.08$. A major difference between the two works is that they consider a highly spinning BH ($a_* = 0.9375$) while we have considered a nonspinning BH ($a_* = 0$), and $R_{\rm ISCO}$ (if it is at all relevant) differs by a factor of 3. A detailed investigation of the effect of BH spin is left for future work. However, some of the discrepancy might also be because we use Equation (14) to calculate $\dot{M}_B$.

As a final note, all the other properties of magnetically dominated simulations which we highlighted earlier —$\rho \propto r^{-1}$, $\beta \sim$ unity, $\eta \approx 2\%$, $\phi_b \approx 30$–40—remain valid for the wide range of simulations ($R_B = 10^2$–$10^7\, r_g$) we have considered here.

## 7. Rotation in the External Gas

In Section 5.3, we found the intriguing result that even when the gas outside the Bondi radius is initialized with coherent rotation, the accreting gas quickly develops random episodes of counterrotation such that, in the final state, there is effectively no coherent rotation at any radius. An immediate suspicion is that this result is an artifact introduced by our multizone method, specifically the Dirichlet radial boundary conditions at annulus boundaries. Here we test this explanation using simulations with a smaller Bondi radius, $R_B \approx 400\, r_g$. A more detailed investigation of the physics of angular velocity loss in strongly magnetized accretion is left for the future (B. S. Prather et al. 2024, in preparation).

### 7.1. Small-scale Tests

Similar to the test presented in Section 3, we once again consider a smaller-scale problem, $R_B \approx 400\, r_g$, and run separate simulations, one with four zones and one with one zone. The setup is identical to that of the strongly magnetized simulation ($\beta_{\rm init} \sim 1$) in Section 3.2. However, instead of initializing the gas with zero velocity, here we initialize it with a strong rotation, $u^\varphi = 0.9\, r^{-3/2} \sin\theta$, with other velocity components still kept zero ($u^r = u^\theta = 0$).

The time evolution of the $\theta$-, $\varphi$-averaged angular velocity $\Omega \equiv u^\varphi/u^t$ at a fixed radius $r = 10\, r_g$, $\Omega_{10}$, in the two simulations is shown in Figure 8. In both the one-zone and four-zone simulations, the initially strong rotation immediately diminishes and the sign of $\Omega_{10}$ keeps flipping throughout the rest of the simulation. This is very similar to what we found in the GIZMO MHD simulation in Section 5.3. The main difference is that there the initial rotation was weak, $\Omega \sim 0.1\, \Omega_K$, whereas here it is strong, $\Omega \sim 0.9\, \Omega_K$. Importantly, the one-zone simulation, which has no internal (Dirichlet) boundaries, displays exactly the same behavior as the four-zone simulation.

Since $\Omega$ keeps switching signs, it is not useful to compare the profiles of $\langle \Omega(r) \rangle$ of the two simulations. Instead, in Figure 9, we compare the $t$, $\theta$, $\varphi$ average of the absolute angular velocity $|\Omega|$ over $\Omega_K$. The agreement in the radial profile of this quantity between the one-zone and four-zone simulations is quite good. We thus conclude that the unusual behavior of angular velocity in these simulations is not an artifact of the Dirichlet boundary condition used in the multizone method.

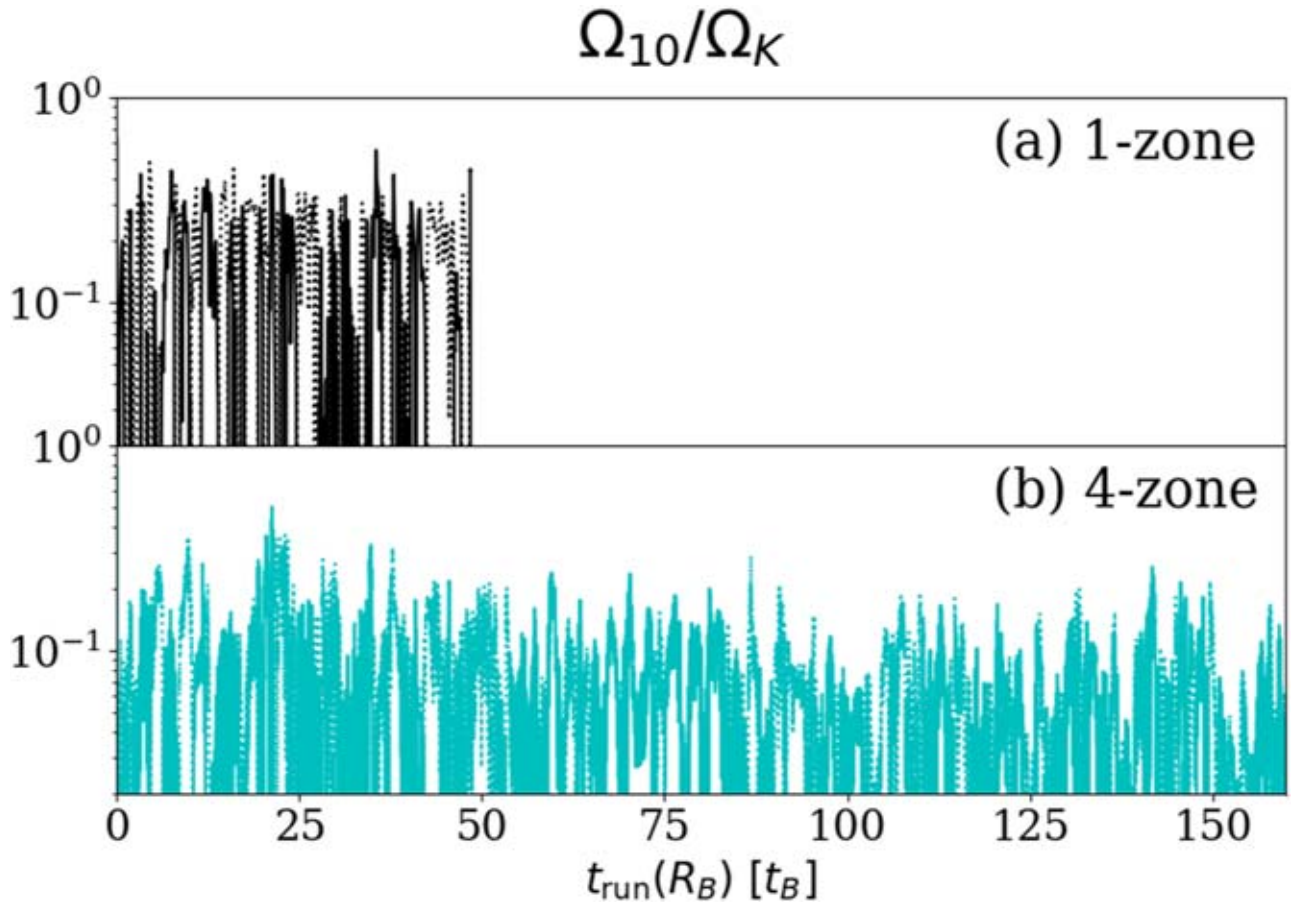

**Figure 8.** Time evolution of the $\theta$-, $\varphi$-averaged angular velocity, $\Omega_{10}$, at $r = 10\, r_g$, over the Keplerian angular velocity $\Omega_K$, for the (a) one-zone and (b) four-zone simulations. Solid and dotted lines correspond to episodes of positive and negative $\Omega_{10}$, respectively.

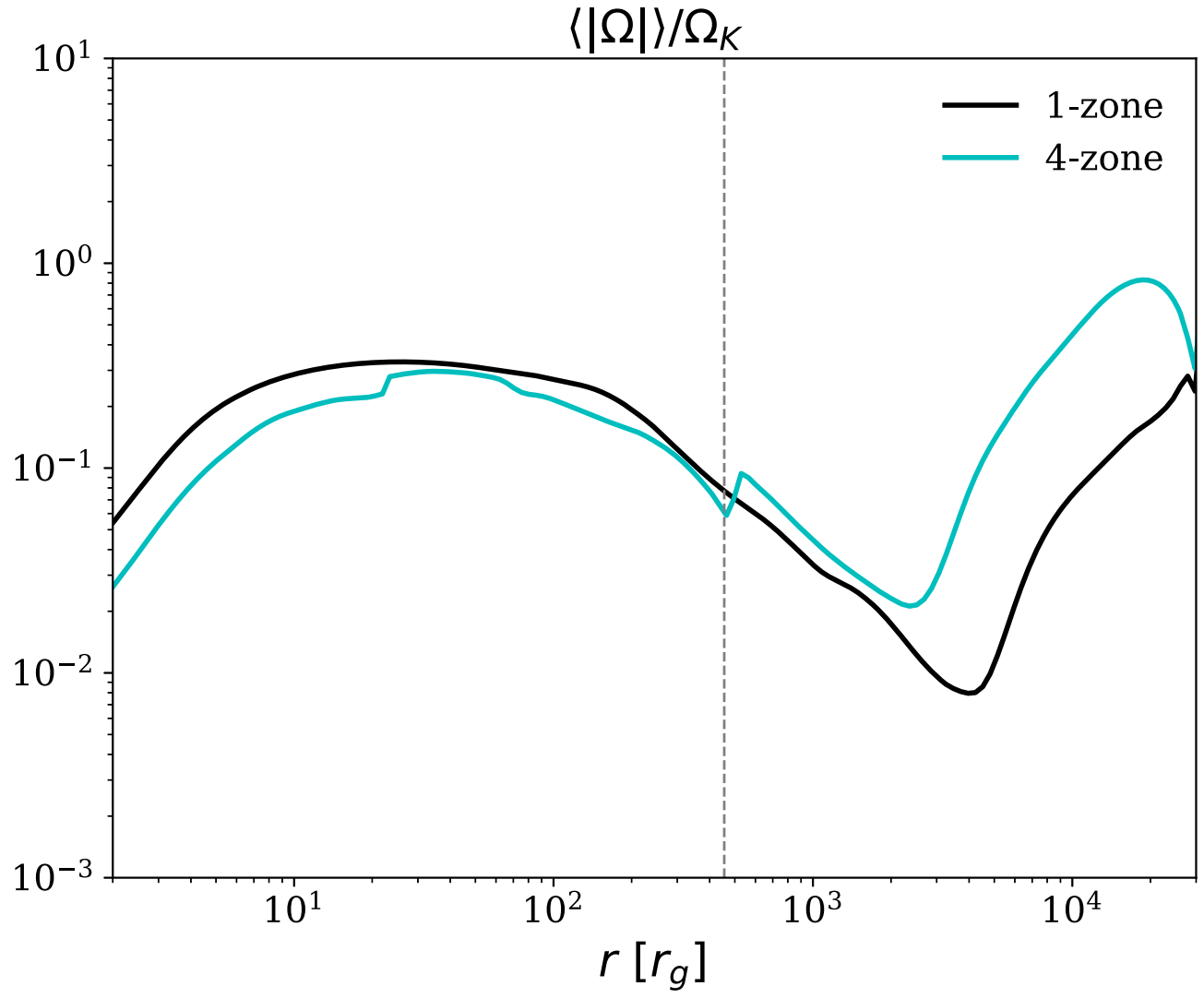

**Figure 9.** Comparison of the converged $t$-, $\theta$-, $\varphi$-averaged absolute angular velocity $\langle|\Omega|\rangle$ over the Keplerian angular velocity $\Omega_K$ between the one-zone (black) and four-zone (cyan) simulations. The two profiles are broadly in agreement.

There is still an unresolved question of why the high level of rotation that is present in the initial conditions of these test models is not maintained in the final steady state. This is quite unlike the standard GRMHD simulations of BH accretion in the literature, where the gas maintains a high level of initial rotation. The majority of those simulations use the L. G. Fishbone & V. Moncrief (1976) torus solution as initial conditions, except for a handful that initialize differently (e.g., S. M. Ressler et al. 2020, 2021; A. Lalakos et al. 2022; N. Kaaz et al. 2023). Our initial conditions are significantly different from typical torus runs. The gas is extended to much larger distances ($r > 10^3\, r_g$ even in our "small-scale" tests, whereas most tori have pressure maxima well inside $10^2\, r_g$), our initial $\beta$ is of order unity (most torus runs use $\beta \sim 10^2$), and our simulations have been run for much longer than previous simulations. These factors could all contribute at some level to cause the difference in the results. We note that U.-L. Pen et al. (2003) also found "magnetic braking" of switching rotation direction in their Newtonian MHD simulation when gas is supplied from a large radius, and R. Narayan et al. (2012) found a steady decrease of $\Omega/\Omega_K$ with time in their long-duration GRMHD MAD simulation.

Another potentially important difference between our work and most other GRMHD simulations discussed in the literature is that all our simulations are performed for the case of a nonspinning BH ($a_* = 0$), which means that the spacetime near the BH does not have any special axis or sense of rotation. When gas flows in from a large radius, it receives no guidance from the BH of the existence of a preferred rotation axis. A spinning BH could be very different as it could, in principle, dictate the rotational dynamics of the accreting gas through frame-dragging, at least at small radii. This question will be explored in B. S. Prather et al. (2024, in preparation).

Although the above test comparing the four-zone and one-zone simulations clearly shows that the unusual behavior of rotation is not a consequence of using the multizone method, we have carried out one more test. In addition to the fiducial `bflux0` prescription that we used in the above four-zone simulation, which does not permit magnetic field lines to move relative to the Dirichlet radial boundaries, we explore another boundary condition, `bflux-const`, that permits a mean rotation at annulus boundaries (Appendix E.2). We have run a four-zone simulation with the `bflux-const` boundary condition, and the results are very similar to what we find with `bflux0`. We anticipate that, when we simulate accretion on a spinning BH, the gas will maintain its rotation over a range of radii, and we will then need to use the `bflux-const` prescription to obtain consistent results.

## 8. Summary and Conclusion

In this paper, we have presented details of a multizone method introduced in H. Cho et al. (2023) to "bridge scales" across 8 orders of magnitude in radius. The multizone method has been implemented in the GRMHD code KHARMA and is designed to tackle, from first principles, the AGN-feeding galaxy-feedback problem, which requires the ability to simulate scales extending all the way from the event horizon of the central SMBH to galactic scales.

The multizone method directly seeks the quasi-steady-state solution for a given large dynamic range system, without getting bogged down by trying to simulate rapidly evolving regions of the system in excessive detail. For the specific purpose of finding an approximate steady state, the method is very efficient. However, it is incapable of providing reliable information on the detailed short-timescale variability characteristics of the system. Traditional simulation techniques operating on a single grid are best suited for the latter.

In Section 3, we presented tests of the method by simulating the accretion of magnetized plasma on a nonspinning BH for an artificially small Bondi radius, $R_B \approx 400\, r_g$. The virtue of the small problem size is that it can be run using standard GRMHD techniques on a single zone. We treat the one-zone simulation as the "true" solution and compare it with the solution we obtain with four zones. We thereby demonstrate that the multizone method produces consistent results. This is true both with ultraweak magnetic fields where field lines are advected with the hydrodynamic flow and exert no back-reaction (Section 3.1), and with strong magnetic fields (plasma-$\beta \approx 1$) where the field dominates the dynamics of the accreting gas and modifies the accretion flow drastically relative to the hydrodynamic Bondi problem (Section 3.2). These tests confirm the validity of the multizone method for bridging scales.

We then extended the study to GRMHD accretion with a realistic $R_B \approx 2\times10^5\,r_g$, similar to the problem studied in H. Cho et al. (2023), but now including realistic boundary conditions (density, temperature, velocity, magnetic field) outside the Bondi radius derived from a galaxy-scale GIZMO simulation, and including the gravitational potential of the galaxy in the GRMHD spacetime metric. Both a GRHD simulation with no magnetic fields (Section 4) and a GRMHD simulation with strong magnetic fields (Section 5) give solutions at radii below the Bondi radius that are remarkably similar to those obtained previously (H. Cho et al. 2023) using a uniform and stationary external medium. This implies that there is a well-defined converged state for accretion on a nonspinning BH, especially strongly magnetized accretion in a magnetically arrested configuration, that is insensitive to the precise details of boundary conditions outside $R_B$. The solutions outside the Bondi radius do depend on the boundary conditions there, as expected.

Coming to specifics, we find that the gas density inside $R_B$ scales with radius as $\rho\propto r^{-1}$, the mass accretion rate on the BH $\dot{M}$ is roughly 1% of the Bondi accretion rate $\dot{M}_B$ (this is for $R_B\approx 2\times10^5 r_g$), and the plasma-$\beta$ is of order unity throughout the volume inside the Bondi radius, falling to yet smaller values near the horizon (corresponding to a magnetization parameter $\phi_b(r_H)\sim 30$), indicating that we have an MAD state not only at the horizon but throughout the Bondi volume. Most significantly, we find positive energy feedback from the accretion flow to the external medium at a level of 2% of $\dot{M}c^2$ in the magnetized accretion problem, independent of the problem setup. The energy penetrates well outside the Bondi radius, up to $10^2 R_B$ in our simulations, and appears to be a robust feature of magnetized accretion on a nonspinning BH.

In Section 6, we described seven multizone GRMHD simulations covering a wide range of Bondi radii, $R_B = 10^2{-}10^7\,r_g$. There is a very clear evolution of the accretion rate with $R_B$: $\dot{M}/\dot{M}_B\approx (R_B/6r_g)^{-0.5}$. Since gas conditions in galactic nuclei vary from one galaxy to another, $R_B/r_g$ is also expected to vary, so the scaling relation we have derived could be used to obtain a better estimate of the actual mass accretion rate on an SMBH in a given galaxy. Apart from the dependence of $\dot{M}/\dot{M}_B$ on the Bondi radius, all the other properties described in the previous paragraph, viz., $\rho\propto r^{-1}$, $\beta\sim$ unity, $\phi_b\approx 30{-}40$, $\eta\approx 2\%$, remain unchanged across a very wide range of Bondi radii.

In Section 7, we reported an unexpected result with regard to the rotation of the accreting gas. When the simulated system extends over many decades of radius and the accreting gas is strongly magnetized, we find that any initial rotation in the gas has a negligible effect on the accretion flow. As a result, the steady flow solution near the BH is virtually independent of the initial rotation. We show that this result is not an artifact of our multizone setup.

In future work, we plan to use the multizone method to carry out GRMHD simulations of magnetized accretion on spinning BHs, again covering up to 8 orders of magnitude in radius. Such a study will reveal which of the results in the present paper carry over to spinning BHs. It will also enable us to focus on relativistic jets and winds launched by a spinning BH, whose feedback efficiency is expected to be far larger than the 2% reported here ($\eta$ can be 100% or more for a rapidly spinning BH, A. Tchekhovskoy et al. 2011; R. Narayan et al. 2022), and whose impact on galactic scales can be explored using multiple zones. The long-term goal of this project is to derive realistic prescriptions for BH accretion and feedback that could serve as subgrid models in galactic/cosmological scale simulations.

## Acknowledgments

We thank the referee for constructive feedback and suggestions to improve the manuscript. H.C., K.S., R.N., and P.N. were partially supported by the black hole Initiative at Harvard University, which is funded by the Gordon and Betty Moore Foundation grant 8273, and the John Templeton Foundation grant 61497. The opinions expressed in this publication are those of the authors and do not necessarily reflect the views of the Moore or Templeton Foundation. This work used Delta at the University of Illinois at Urbana-Champaign through allocations PHY230079 and AST080028 from the Advanced Cyberinfrastructure Coordination Ecosystem: Services & Support (ACCESS) program, which is supported by National Science Foundation grants #2138259, #2138286, #2138307, #2137603, and #2138296. This research used resources provided to B.P. by the Los Alamos National Laboratory Institutional Computing Program, which is supported by the U.S. Department of Energy National Nuclear Security Administration under Contract No. 89233218CNA000001.

## Appendix A
## Calculating the Analytic Relativistic Bondi Solution

The simplest accretion model is the classical H. Bondi (1952) accretion under Newtonian gravity. In this adiabatic, purely hydrodynamic, spherically symmetric accretion flow, the gas radially infalls to the central object and the BH's asymptotic velocity relative to the ambient gas is zero $v(\infty)=0$. In this Newtonian solution, the Bondi accretion rate is

$$\dot{M}_B \equiv 4\pi r_s^2\rho(r_s)c_s(r_s) = \pi G^2 M_\bullet^2\frac{\rho_\infty}{c_{s,\infty}^3}\left[\frac{2}{5-3\gamma}\right]^{(5-3\gamma)/2(\gamma-1)}, \tag{A1}$$

where $r_s$ is the sonic radius and the last factor becomes 4.5 in the limit of $\gamma=5/3$ (J. Frank et al. 2002).

F. C. Michel (1972, see also S. L. Shapiro & S. A. Teukolsky 1983) later followed up by deriving a general relativistic solution for spherical accretion on a Schwarzschild black hole ($a_*=0$) where the solution is parameterized by the sonic radius $r_s$. Here we will work in units of $GM_\bullet = c = 1$. There are two constant quantities, the mass flux $C_1\equiv\rho u^r r^2$ and the Bernoulli parameter $C_2\equiv (T^r_t/(\rho u^r))^2$. The second constant can be expressed as

$$C_2 = \left(\frac{T_t^r}{\rho u^r}\right)^2 = [1+(1+n)T]^2\left(1-\frac{2}{r}+(u^r)^2\right), \tag{A2}$$

where $n = 1/(\gamma_{\rm ad}-1)$. At the sonic point, the temperature $T_s = T(r_s)$ and the radial velocity $u_s^r = -(2r_s)^{-1/2}$ can be expressed as functions of the sonic radius $r_s$. Therefore equating $C_2$ at the sonic point $C_2(r_s)$ and at infinity $C_2(\infty) = [1+(1+n)T_\infty]^2$ gives an approximate relation between the Bondi radius $R_B = 1/(\gamma_{\rm ad} T_\infty)$ and the sonic radius $r_s$. Assuming $u_s^r$, $T_\infty \ll 1$, the approximate formula for $R_B(r_s)$ is

$$R_B = \frac{1}{\gamma_{\rm ad} T_\infty} \approx \begin{cases} \frac{4(1+n)}{2(n+3)-9}\, r_s & \propto r_s \quad (\gamma_{\rm ad} \neq 5/3), \\ \frac{80}{\gamma_{\rm ad} 27} r_s^2 & \propto r_s^2 \quad (\gamma_{\rm ad} = 5/3). \end{cases} \tag{A3}$$

The Bondi solutions are obtained numerically by solving the equation

$$f(T) = [1 + (1+n)T]^2 \left(1 - \frac{2}{r} + \left(\frac{C_1}{T^n r^2}\right)^2\right) - C_2 = 0, \tag{A4}$$

where $u^r$ in Equation (A2) has been rewritten in terms of $C_1$. It is useful to know the initial guess of the temperature at a given radius. At large radii, $T_\infty = (\sqrt{C_2} - 1)/(1+n)$, and at small radii, assuming a freefall velocity $u^r = (2r)^{-1/2}$, $T_{\rm near} \approx (C_1/\sqrt{2/r^3})^{1/n}$.

## Appendix B
## Including External Gravity in GRMHD

The KHARMA code that we employ for our multizone GRMHD simulations includes the Kerr metric as a built-in default for BH accretion simulations. The spin-0 limit of Kerr, the Schwarzschild metric, applies to the simulations in this paper. However, when we run simulations that include boundary conditions from an external galaxy, we need to generalize the metric to include the gravitational effect of the external galaxy. This problem is discussed here.

In the weak-field Newtonian limit, the general relativistic spacetime of a spherically symmetric object simplifies to

$$ds^2 = -(1 + 2\Phi)dt^2 + dr^2 + r^2 d\theta^2 + r^2 \sin^2\theta d\varphi^2, \tag{B1}$$

where $\Phi(r)$ is the Newtonian gravitational potential, which in our case is the sum of the gravitational potentials of the BH and the external galaxy (assumed to be spherically symmetric): $\Phi(r) = -(1/r) + \Phi_g(r)$. Guided by this limit, we modify the strong-field Schwarzschild metric of a nonspinning BH to the following spherically symmetric metric, which includes the potential of the galaxy:

$$ds^2 = -\left(1 - \frac{2}{r} + 2\Phi_g\right)dt^2 + \left(1 - \frac{2}{r} + 2\Phi_g\right)^{-1} dr^2 + r^2 d\theta^2 + r^2 \sin^2\theta d\varphi^2. \tag{B2}$$

This metric smoothly interpolates between small radii where the BH's gravity dominates the spacetime and large radii where the galaxy dominates. Technically, we need only modify $g_{tt}$ since that is the only component of the metric that contributes to the Newtonian limit. However, we make a similar modification also in $g_{rr}$. This change is harmless in the regions where the galaxy dominates, but it has the virtue of preserving the BH character of the metric at small radii. In particular, the horizon radius $r_H$ of the metric (B2) is determined by the condition

$$1 - \frac{2}{r_H} + 2\Phi_g(r_H) = 0. \tag{B3}$$

Since $\Phi_g(r_H)$ is expected by $\ll$1, the new $r_H$ will be very close to the Schwarzschild radius $2\, r_g$. Furthermore, as we are free to choose the zero-point of the galaxy potential, we could avoid even this small shift in the location of the horizon by choosing to set $\Phi_g(2\, r_g) = 0$. With this choice, the horizon will be located precisely at $r = 2\, r_g$, just as in the original Schwarzschild metric. This is the approach we have taken in our model of the galaxy potential in Equation (13).

For running GRMHD simulations we need a horizon-penetrating version of the spacetime described by Equation (B2). Following standard procedures used to derive the Eddington–Finkelstein metric, we find

$$ds^2 = -\left(1 - \frac{2}{r} + 2\Phi_g\right)dt^2 + 4\left(\frac{1}{r} - \Phi_g\right)dt\, dr + \left(1 + \frac{2}{r} - 2\Phi_g\right)dr^2 + r^2 d\theta^2 + r^2 \sin^2\theta d\varphi^2. \tag{B4}$$

This generalization of the Eddington–Finkelstein metric includes the effect of the galaxy at large radii and is guaranteed to be well-behaved at the horizon.

The metric (B4) is spherically symmetric and is appropriate for a nonspinning BH at the center of a galaxy, which is the case for all the simulations described in this paper. For a spinning BH embedded in a (mostly spherical) galaxy, Equation (B4) can be modified following the methods described in P. Kocherlakota et al. (2023, inspired by E. T. Newman & A. I. Janis 1965; Azreg-Aïnou 2014) to generate the appropriate metric.

## Appendix C
## Coordinate Systems

There are three types of coordinate systems used in this work: eKS, MKS, and WKS. The code coordinates $x^r$, $x^\theta$, $x^\varphi$ are spaced evenly.

In eKS coordinates, $x^r = \log r$, $x^\theta = \theta$, $x^\varphi = \varphi$.

The MKS (C. F. Gammie et al. 2003) is the same as eKS but with a modified $\theta$ grid,

$$\theta = \pi x^\theta + \frac{1}{2}(1 - h)\sin(2\pi x^\theta), \tag{C1}$$

where $h = 0.3$ and $x^\theta \in [0, 1]$. This results in focusing the resolution near the midplane.

Finally, we propose a new type of coordinate system, the WKS. In WKS, the resolution near the poles is coarse and there is a nearly constant resolution elsewhere. While the radial and azimuthal grid spacings remain the same as in eKS and MKS, the $\theta$ grid is modified to

$$\theta = \frac{\pi}{2}\left[1 + f_{\rm lin}(2x^\theta - 1) + (1 - f_{\rm lin}) \times \left\{\tanh\left(\frac{x^\theta - 1}{\lambda}\right) + 1\right\} - (1 - f_{\rm lin})\left\{\tanh\left(-\frac{x^\theta}{\lambda}\right) + 1\right\}\right], \tag{C2}$$

where $f_{\rm lin} = 0.6$ is the linear fraction where the $\theta$ grid is spaced evenly. The rest $(1 - f_{\rm lin})$ has a nonlinear spacing following tanh(), which results in wider cells near the poles. $\lambda = 0.03$–$0.04$ is the smoothness parameter that determines how smooth the transition is between the linear and nonlinear regions. $\theta$ as a function of $x^\theta$ is shown by the blue solid line in Figure 10.

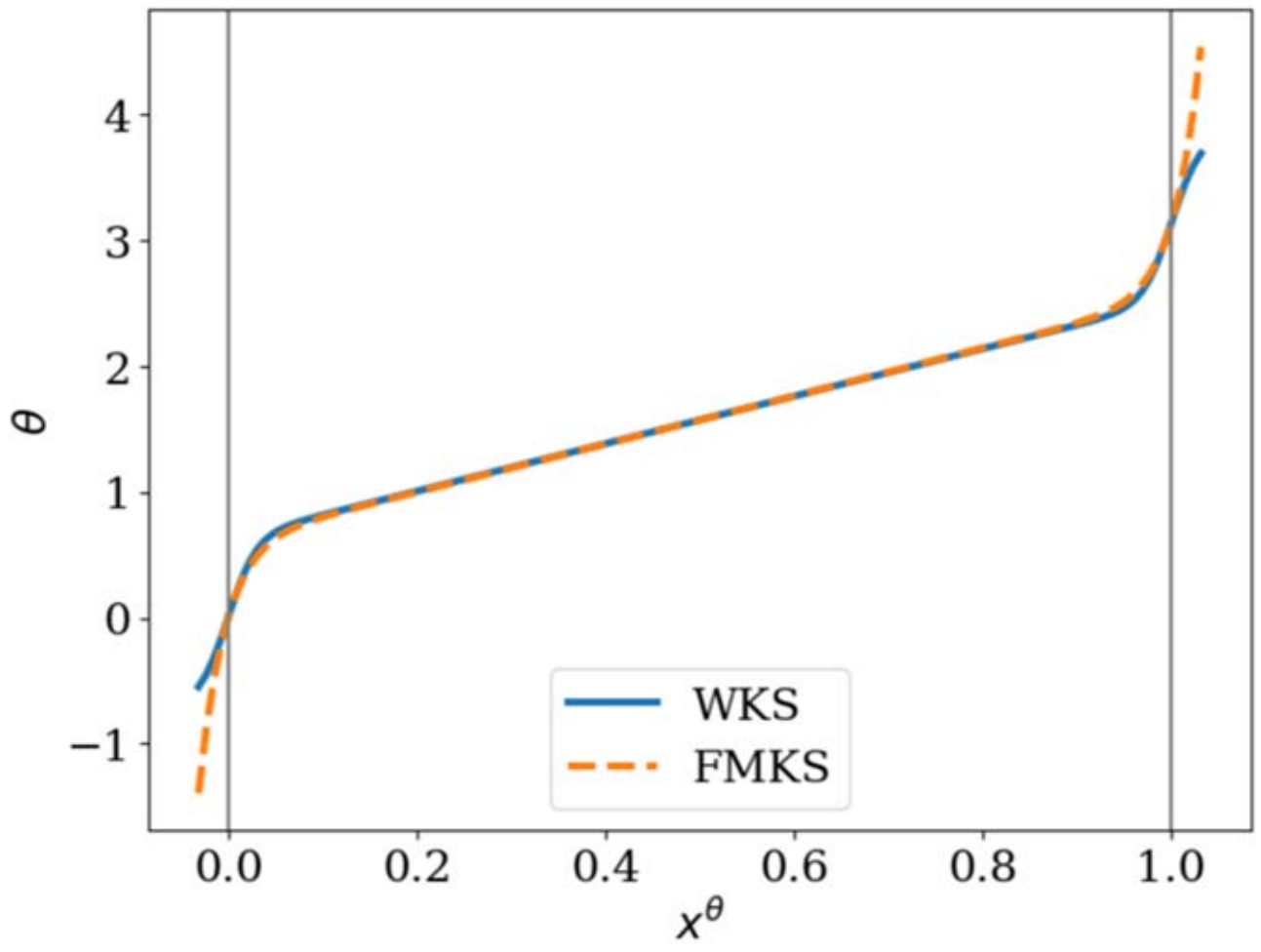

**Figure 10.** $\theta$ grid in our proposed WKS coordinate system (solid blue line) compared with the widely used FMKS coordinate system ($\chi_t = 0.87$, $\alpha = 18$; dashed orange line), both shown as functions of the code coordinate $x^\theta$. The physical zones are $x^\theta \in [0, 1]$ between the two vertical gray lines. The FMKS and WKS coordinate systems are almost identical in the physical zone $0 \leqslant x^\theta \leqslant 1$ but differ in the ghost zones beyond the two poles.

The new coordinate system WKS is effectively similar to the funky modified Kerr–Schild (FMKS) coordinate system which is widely used in GRMHD simulations, including our earlier work (H. Cho et al. 2023). The FMKS coordinates without cylindrification have a $\theta$ grid given by

$$\theta = Ny\left(1 + \frac{(y/\chi_t)^\alpha}{\alpha + 1}\right) + \frac{\pi}{2}, \tag{C3}$$

where $N \equiv \pi/2(1 + \chi_t^{-\alpha}/(1 + \alpha))^{-1}$ is a normalization factor, $y = 2x^\theta - 1$, and the typical choices for the parameters are $\chi_t = 0.82$, and $\alpha = 14$. In Figure 10, the $\theta$ grid as a function of the code coordinate $x^\theta$ is shown for the WKS (blue solid) and FMKS (orange dashed) coordinate systems. FMKS with $\chi_t = 0.87$ and $\alpha = 18$ closely follows the WKS coordinates with $f_{\rm lin} = 0.6$, $\lambda = 0.03$ over the physical zone, $x^\theta \in [0, 1]$. In the ghost zones ($x^\theta < 0$, $x^\theta > 1$), the WKS $\theta$ grid size is symmetric with respect to the poles due to the tanh() function while the FMKS cells get larger in the $\theta$-direction. Therefore, one benefit of using WKS over FMKS is that the ghost cells are reasonably sized. The other benefit of WKS is that it is relatively easy to understand intuitively compared to FMKS, so the parameters can be chosen based on the desired fraction of the $\theta$ range that should have uniform resolution and the desired smoothness of the transition between the linear and nonlinear regions of $\theta$.

## Appendix D Initialization of the Magnetic Field

All simulations are initialized with a constant plasma-$\beta$ profile across the wide dynamic range in radius. In order to do so, the magnetic field initialization depends on the initial gas pressure. For a given $\rho$ and $T$ radial scaling, if the gas pressure $p_g = \rho T \propto r^p$, a vector potential with $A_\varphi \propto r^{p/2+1} \sin\theta$ as the only nonzero component generates a nearly vertical magnetic field with an approximately constant $\beta$ profile.

The MH. Bondi runs in H. Cho et al. (2023) had initial conditions of $\rho \propto r^{-1}$, $T \propto r^{-1}$ at small radii $r \ll R_B$ and $\rho$, $T \propto r^0$ constant at large radii $r \gg R_B$. Therefore, the vector potential in KHARMA was initialized via Equation (7) such that $A_\phi \propto \sin\theta$ at $r \ll R_B$ and $A_\phi \propto r \sin\theta$ at $r \gg R_B$.

For the runs that are initialized from GIZMO data in Section 5, the initial magnetic field in the galaxy simulation has both toroidal $\boldsymbol{B}_{\rm tor}$ and poloidal $\boldsymbol{B}_{\rm pol}$ components. The toroidal component $\boldsymbol{B}_{\rm tor}$ of $\beta \sim 10^3$ dominates at large radii $r > r_0$ and poloidal field $\boldsymbol{B}_{\rm pol}$ of $\beta \sim 10$ dominates at small radii $r < r_0$ where $r_0 = 10$ kpc. Since the initial condition is $\rho \propto r^{-1}$ and $T$ is roughly constant for 300 pc $< r <$ 1 Mpc, the vector potential of the poloidal magnetic component is set to be

$$A_{\varphi,\rm pol} = \frac{b_z}{2} r^{1/2} \sin\theta e^{-r/r_0}, \tag{D1}$$

which produces a divergence-free vertical field of constant $\beta$ interior to $r_0$. Outside $r_0$, the strength of the poloidal field $\boldsymbol{B}_{\rm pol}$ drops quickly and the toroidal field $B_{\rm tor}$ takes over. To set up the initial toroidal component of the magnetic field, we calculate the magnetic field strength for each position from the thermal energy density with a given $\beta \sim 1000$ and assign the magnetic field to $B_\phi$.[15]

## Appendix E Boundary Conditions for the Magnetic Field

The Dirichlet radial boundary condition holds magnetic fields in the ghost zones fixed, while the magnetic fields in the physical zones evolve following flux-CT. Since ghost zones are not updated consistent with the magnetic field fluxes, without special treatment the divergence at the interface between the ghost and the physical zones will increase. Here we describe two types of prescriptions at Dirichlet boundaries to keep the magnetic fields divergence-free. The first prescription `bflux0` is used in all runs in H. Cho et al. (2023) and in this work. This simpler prescription keeps the effective flux of magnetic fields through the radial boundaries zero and thus is consistent with the Dirichlet boundary condition. The other prescription, `bflux-const`, is a generalized version, which better handles systems with coherent rotation. This version of the prescription is not used for the runs in this work since `bflux0` already gives consistent results, but it will be adopted for future simulations with spinning BHs where a high level of rotation is anticipated. The implementation details of `bflux0` and `bflux-const` are given in Appendices E.1 and E.2, respectively, and the effect of the two prescriptions on a simulation is described in Appendix E.3.

### E.1. `bflux0` Prescription

For simplicity, we will explain the prescription in two dimensions first and then generalize to three dimensions. The superscript $n$ indicates the timestamp and subscripts indicate the spatial location in the grid with integer subscripts $i, j, k \in \mathbb{Z}$ are corresponding to cell centers.

The divergence at the corner $(i_0 + 1/2,\ j_0 + 1/2)$ in the interface between the physical and ghost cells (shown in a pink

[15] The poloidal component of the initial magnetic field in the GIZMO simulation is strictly divergence-free. Initially, the thermal pressure profile is roughly axisymmetric, so the divergence of the toroidal component is ensured to be small. After the start of the simulation, the divergence cleaning in GIZMO further damps the divergence.

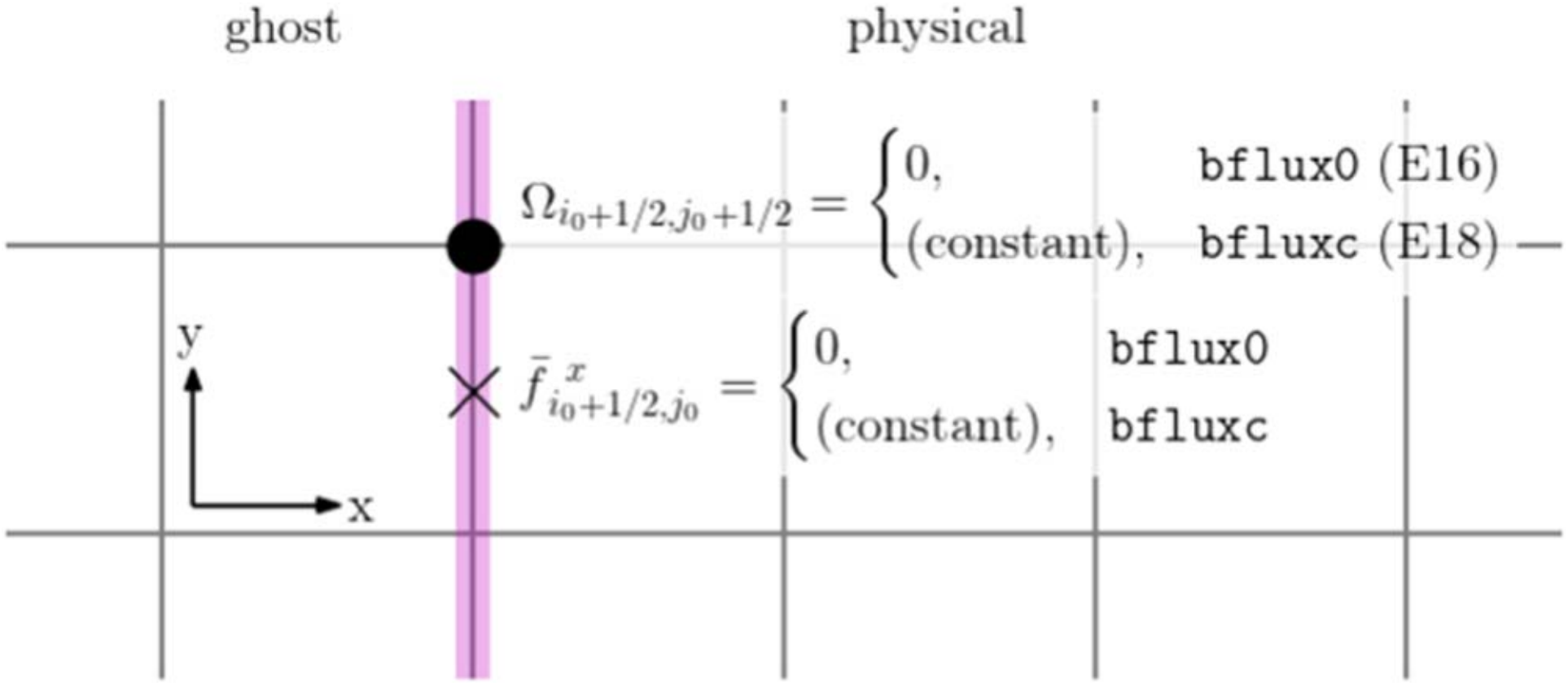

**Figure 11.** Comparison between two prescriptions for the magnetic fields at the radial Dirichlet boundaries. The `bflux0` prescription sets the electric field $\Omega$ to 0 at the boundary, and the `bflux-const` prescription averages the electric field $\Omega$ over all boundary cells in the $y$-direction and replaces the individual $\Omega$ values with this average. This results in `bflux0` having a zero flux of magnetic field across the boundary, $\bar{f}^x = 0$, and `bflux-const` having a constant mean flux of magnetic field to better handle the mean $y$-motion of the gas.

line in Figure 11 is

$$(\nabla \cdot B)^{n+1}_{i_0+1/2,j_0+1/2} = \frac{1}{2\Delta x}\left(B^{x,n+1}_{i_0+1,j_0} + B^{x,n+1}_{i_0+1,j_0+1} - \boldsymbol{B}^{\mathbf{x,n+1}}_{\mathbf{i_0,j_0}} - \boldsymbol{B}^{\mathbf{x,n+1}}_{\mathbf{i_0,j_0+1}}\right) + \frac{1}{2\Delta y}\left(\boldsymbol{B}^{\mathbf{y,n+1}}_{\mathbf{i_0,j_0+1}} + B^{y,n+1}_{i_0+1,j_0+1} - \boldsymbol{B}^{\mathbf{y,n+1}}_{\mathbf{i_0,j_0}} - B^{y,n+1}_{i_0+1,j_0}\right).$$

In Dirichlet boundary conditions, the ghost cell magnetic fields, marked in bold above, stay constant: $B^{l,n+1}_{i_0,j} = B^{l,n}_{i_0,j} (l \in \{x, y\}, \forall j)$ stays constant. The physical cells are evolved using the averaged fluxes $\bar{f}$ following flux-CT $B^{x,n+1}_{i,j} = B^{x,n}_{i,j} - \Delta t(\bar{f}^y_{i,j+1/2} - \bar{f}^y_{i,j-1/2})/\Delta y$, $B^{y,n+1}_{i,j} = B^{y,n}_{i,j} - \Delta t(\bar{f}^x_{i+1/2,j} - \bar{f}^x_{i-1/2,j})/\Delta x$. Therefore, substituting all the $B^{l,n+1}_{i_0+1,j}$ ($i = i_0 + 1$ black terms) with the above rules and replacing $B^{l,n+1}_{i_0,j}$ ($i = i_0$ bold terms) with $B^{l,n}_{i_0,j}$, the divergence equation becomes

$$\begin{aligned}(\nabla \cdot B)^{n+1}_{i_0+1/2,j_0+1/2} = (\nabla \cdot B)^{n}_{i_0+1/2,j_0+1/2} &- \frac{\Delta t}{2\Delta x\Delta y}\left(\cancel{\bar{f}^y_{i_0+1,j_0+1/2}} - \bar{f}^y_{i_0+1,j_0-1/2}\right. \\ &\left. + \bar{f}^y_{i_0+1,j_0+3/2} - \cancel{\bar{f}^y_{i_0+1,j_0+1/2}}\right) \\ &- \frac{\Delta t}{2\Delta x\Delta y}\left(\bar{f}^x_{i_0+3/2,j_0+1} - \bar{f}^x_{i_0+1/2,j_0+1}\right. \\ &\left. - \bar{f}^x_{i_0+3/2,j_0} + \bar{f}^x_{i_0+1/2,j_0}\right). \end{aligned} \tag{E1}$$

The six barred fluxes $\bar{f}$ are not fully canceled out because the ghost and physical cells are evolved differently. The averaged fluxes are

$$\bar{f}^x_{i+1/2,j} = \frac{1}{2}(\Omega_{i+1/2,j-1/2} + \Omega_{i+1/2,j+1/2}), \tag{E2}$$

$$\bar{f}^y_{i,j+1/2} = -\frac{1}{2}(\Omega_{i-1/2,j+1/2} + \Omega_{i+1/2,j+1/2}), \tag{E3}$$

where $\Omega$ is the $z$-component of the electric field $E^z$ (EMF) at the corners (calculated following D. S. Balsara & D. S. Spicer 1999). In the `bflux0` prescription, the EMFs are set to zero,

$$\Omega_{i_0+1/2,j+1/2} = 0\ (\forall j), \tag{E4}$$

at the interface. This results in the averaged flux across the interface also being zero, $\bar{f}^x_{i_0+1/2,j} = 0\ (\forall j)$, which removes two averaged fluxes $\bar{f}$ in Equation (E1). The remaining four averaged fluxes also cancel out with the same Equation (E4) and the divergence is conserved at the boundary $(\nabla \cdot B)^{n+1}_{i_0+1/2,j_0+1/2} = (\nabla \cdot B)^{n}_{i_0+1/2,j_0+1/2}$.

The prescription can be generalized to three dimensions by putting the $y$- and $z$-components of the electric field to zero:

$$\Omega^y_{i_0+1/2,j,k+1/2} = \Omega^z_{i_0+1/2,j+1/2,k} = 0 \qquad (\forall j, k). \tag{E5}$$

In our case, the $x$-, $y$-, $z$-directions correspond to the $r$-, $\theta$-, $\phi$-directions.

### E.2. `bflux-const` Prescription

The above `bflux0` prescription can be generalized where instead of a zero averaged flux $\bar{f}$ across the boundaries, $\bar{f}$ is set to a constant value, hence the name `bflux-const`.

In particular, we choose for the constant the average EMF over the interface,

$$\Omega_{i_0+1/2,j+1/2} = \langle\Omega\rangle_y\ (\text{constant at } \forall j), \tag{E6}$$

where $\langle\Omega\rangle_y$ is the average of all EMFs in the $y$-direction at the $i_0 + 1/2$ boundary.[16] Therefore, the averaged fluxes across the boundary are constant in the $y$-direction because of Equation (E2): $\bar{f}^x_{i_0+1/2,j} = \langle\Omega\rangle_y$ (constant at $\forall j$). One can check that this prescription leads to preserving the divergence $(\nabla \cdot B)^{n+1}_{i_0+1/2,j_0+1/2} = (\nabla \cdot B)^{n}_{i_0+1/2,j_0+1/2}$.

Generalizing to three dimensions requires more caution in the case of `bflux-const`. In spherical coordinates, the $y$-direction corresponds to $\theta$ where the boundary condition is reflecting at the poles, which might conflict with the Dirichlet radial boundary at the $x$, $y$ corners.

When the boundary condition in the $z$-direction ($\varphi$ in spherical coordinates) is periodic, the $y$-component of the EMF

[16] In this appendix subsection, the brackets $\langle\rangle$ are simple averages and are not the density-weighted time averages defined in Equation (8).

can be safely averaged over the $z$-direction at a given $i_0 + 1/2$, $j$

$$\Omega^y_{i_0+1/2,j,k+1/2} = \langle \Omega^y_j \rangle_z. \qquad (\forall j,\, k) \qquad \text{(E7)}$$

Here $\langle \Omega^y_j \rangle_z$ is only a function of $j$ (or $\theta$ in spherical coordinates) at the radial boundary.

When the boundary condition in the $y$-direction is periodic, the $z$-component of the EMF can similarly be averaged. However when it is reflecting, which requires $\Omega^x = \Omega^z = 0$ at the polar boundaries, in order to avoid conflict at the $x$, $y$ boundary corners, we must set $\Omega^z = 0$ at all $\theta$

$$\Omega^z_{i_0+1/2,j+1/2,k} = \begin{cases} \langle \Omega^z_k \rangle_y & (y \text{ periodic}) \\ 0 & (y \text{ reflecting}) \end{cases} (\forall j,\, k). \qquad \text{(E8)}$$

In spherical coordinates, this is effectively equivalent to applying `bflux0` along the $\theta$-direction and `bflux-const` along the $\varphi$-direction. This seems reasonable. It is highly unlikely that we will have a coherent $\theta$-velocity directed toward a single pole, so `bflux0` is okay for the $\theta$-direction. However, it can often happen that the gas flow has coherent rotation in the $\varphi$-direction, and to preserve this rotation we need `bflux-const`.

### *E.3. Comparison Between `bflux0` and `bflux-const` Prescriptions*

In this subsection, we study the effect of the two prescriptions. We first construct a simple two-dimensional problem in Minkowski spacetime in Cartesian coordinates $[0, 1]\, r_g \times [0, 1]\, r_g$ with $16 \times 16$ resolution. The run is initialized with a homogeneous medium of constant density and temperature. A Dirichlet boundary condition is used in the $x$-direction and a periodic boundary condition for the $y$-direction.

For a basic test, the magnetic field is initialized with a uniform pure $x$-component with $\beta \sim 1$, and the velocity is initialized with a uniform pure $y$-component, $\boldsymbol{u} = 0.01c\, \hat{y}$. The simulation is run up to time $t = 100\, t_g$ such that the gas completes one cycle around the periodic boundary. This test is run with both the `bflux0` and `bflux-const` prescriptions and the final snapshots are shown in Figures 12(a) and (b), respectively. Because no flux can cross the boundaries, the

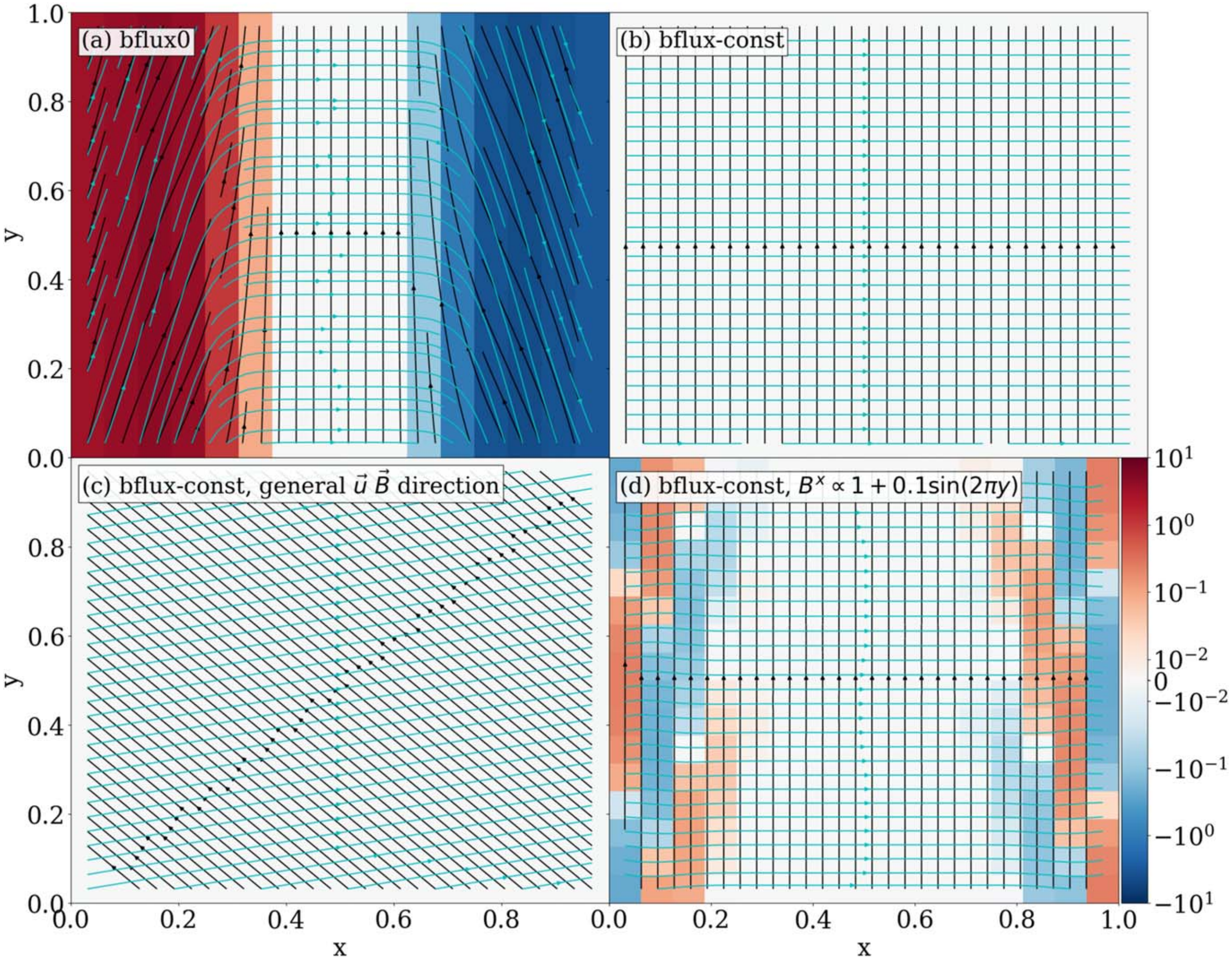

**Figure 12.** The fractional change in $B^y$ after $\Delta t = 100\, t_g$ relative to the initial magnetic field strength $B_0$, $\Delta B^y/B_0$, is shown in the background color for the four tests. The velocity streamline is overlaid in black lines and the magnetic field lines in cyan. The upper panels show the result when the system is initialized with homogeneous $\boldsymbol{u} = 0.01c\, \hat{y}$ and $\boldsymbol{B} \| \hat{x}$ and (a) the `bflux0` prescription is used, or (b) `bflux-const` is used. Using the `bflux0` prescription results in significant deviation from the initial conditions, while `bflux-const` produces no change. (c) The test is repeated for a general angle between the uniform $\boldsymbol{u}$ and $\boldsymbol{B}$ fields with `bflux-const` and the initial state is still maintained. (d) When the initial magnetic field is inhomogeneous, there is a small deviation from the initial state because the `bflux-const` method only translates the mean field. Overall, the $\boldsymbol{u}$ and $\boldsymbol{B}$ fields are less distorted compared to when the `bflux0` prescription is used.

`bflux0` boundary condition drags field lines back at the boundaries even as the fluid advects them in the center. This in turn strongly perturbs the gas flow (as can be seen from the velocity streamlines in black). This is a concern because `bflux0` leaves a strong impact on bulk tangential motion (e.g., coherent rotation). On the other hand, for `bflux-const`, since the field lines are allowed to move with the flow parallel to the boundary, they are able to align with the field in the ghost zones without producing any distortions in the physical zones. Therefore, the initial magnetic field and velocity field are not modified.

Next, a random angle between homogeneous $\boldsymbol{u}$ and $\boldsymbol{B}$ is chosen such as $\boldsymbol{u} = 0.01c(-\hat{x} + \hat{y})$ and $\boldsymbol{B} = B_0(\hat{x} + 0.2\,\hat{y})$. Figure 12(c) shows the result when the `bflux-const` prescription is used. Once again the initial $\boldsymbol{u}$ and $\boldsymbol{B}$ fields are preserved.

Finally, we test `bflux-const` on a case where the magnetic field is not homogeneous. The initial field is chosen to be $\boldsymbol{B} = B_0(1 + 0.1\sin(2\pi y))\hat{x}$ and the velocity is the same as before, $\boldsymbol{u} = 0.01c\,\hat{y}$. In this case, there are some deviations in the magnetic field as shown in the background in Figure 12(d), which can be as large as $\lesssim$10%. The `bflux-const` prescription, through its technique of averaging, successfully transports the mean magnetic field in the $y$-direction, thereby capturing a large fraction of the evolution. However, it cannot handle fluctuations around the mean, and these result in the distortions seen in Figure 12(d). Such perturbations are expected in the magnetic field in any turbulent accretion flow, and an entirely turbulent field (with zero mean $\Omega$) will effectively behave like `bflux0`. However, the ability of `bflux-const` to capture at least the mean field motion is a substantial improvement over `bflux0`.

In summary, the `bflux-const` prescription is more suitable for the case where there is a bulk shear motion with respect to the Dirichlet boundary. In `bflux0` the field lines are anchored to the ghost zone and are dragged with the shear flow which significantly alters the velocity. On the other hand, in `bflux-const`, the mean shear motion is taken into account and only a small impact is made to both $\boldsymbol{u}$ and $\boldsymbol{B}$ fields, so long as $\boldsymbol{B}$ is mostly coherent. As mentioned in Section 7, both prescriptions produce similar results in this work because our simulations do not maintain a high level of rotation. In the future, though, when studying spinning BHs, using `bflux-const` is anticipated to better capture the mean azimuthal motion.

## Appendix F
## Tests of the Multizone Setup

In our earlier work (H. Cho et al. 2023), we showed that the larger-scale simulation ($R_B \approx 2 \times 10^5\,r_g$) converges independent of the resolution, the coordinate system, or the initial conditions. Here we demonstrate a similar test for a smaller-scale simulation ($R_B \approx 400\,r_g$), first focusing on testing the new WKS coordinate system. The fiducial run is chosen to be the one-zone strongly magnetized simulation in Section 3.2. Four extra simulations are run, each varying the mode (single zone versus multizone), resolution, or coordinate system, as summarized in Table 2. Run (i) uses a different coordinate system (MKS) and run (ii) uses a higher resolution of $320 \times 128^2$ compared to the fiducial run. Since the four-zone equivalent of the fiducial run has already been compared and shown to give consistent results in Section 3.2, we omit that run here. Run (iii) is a four-zone equivalent of the higher-resolution run (ii), and run (iv) is similar to run (iii) but in FMKS coordinates. Radial profiles from these runs are compared in Figure 13, with the fiducial run shown with black curves. The agreement among the various runs is very good regardless of whether we use a one-zone or multizone approach, and our choice of resolution or coordinate system. This confirms that our new WKS coordinate system produces reliable results.

**Table 2**
Simulation Setup for Different Runs

| Label | Mode | Resolution Per Zone | Coordinate System |
|---|---|---|---|
| fiducial | one-zone | $160 \times 64^2$ | WKS |
| (i) MKS | one-zone | $160 \times 64^2$ | MKS |
| (ii) 128 | one-zone | $320 \times 128^2$ | WKS |
| (iii) four-zone, 128 | four-zone | $128^3$ | WKS |
| (iv) four-zone, 128, FMKS | four-zone | $128^3$ | FMKS ($\chi_t = 0.8$, $\alpha = 16$) |

Next, we test the effect of varying annulus sizes or different values of the base $b$ for constructing the annulus boundaries. In addition to the fiducial choice of $b = 8$ with $n = 4$ zones, two extra multizone simulations with bases $b = 6$ with $n = 5$ zones and $b = 12$ with $n = 3$ zones were carried out. All other details of the multizone setup are identical for the three simulations and the runtime per zone $t_{\mathrm{MHDcap},i}$ (in Equation (6)) is adjusted consistently for the chosen base as $b^{-3/2}/2\,\min(t_{\mathrm{char},\,i},\,t_B)$. Figure 14 compares the multizone runs with different bases with the one-zone run. While $b = 8$ and $b = 12$ runs converge in all profiles with those of the one-zone simulation at late times, a run with $b = 6$ exhibits a somewhat higher value of feedback efficiency $\eta \approx 2\%$ instead of 1%. For a fixed system size, selecting smaller annuli requires more zones to cover the same radial range and this introduces more internal Dirichlet boundaries. Since the `bflux0` boundary condition results in temporary magnetic tension, more internal boundaries can lead to a higher level of feedback for a small base of $b = 6$. A more serious effect is that using a smaller base puts the innermost internal boundary uncomfortably closer to the black hole at $r = 6\,r_g$ where the accumulated tension is likely to be stronger. Regardless, when compared to one-zone results, we have verified that the fiducial choice of $b = 8$ is still a safe choice where the internal boundaries have a negligible effect.

Finally, we test the dependence on the run duration for each zone. As explained in Section 2.3, an optimal coefficient to multiply to the characteristic timescale $t_{\mathrm{char},i}$ is $\approx$0.04. This corresponds to the characteristic time at the logarithmic center of each zone, where the neighboring zone's boundary is located. Then, this choice of coefficient 0.04 allows enough time to iron out any boundary effects of the previously active zone. Our fiducial choice is half of this factor 0.02 as shown in Equation (6) in order to reduce the level of accumulated boundary effects. Other than the fiducial value, we additionally simulate using different factors (0.01, 0.04, 0.08) around the optimal value and compare with the one-zone run in Figure 15. The runs with factors 0.02, 0.04, and 0.08 all show a remarkable agreement with the one-zone run as expected from the fact that they are close to the suitable choice of 0.04. However, there is a discrepancy in feedback efficiency $\eta$ for the run with a factor of 0.01 where the runtime is 4 times shorter than the optimal factor of 0.04.

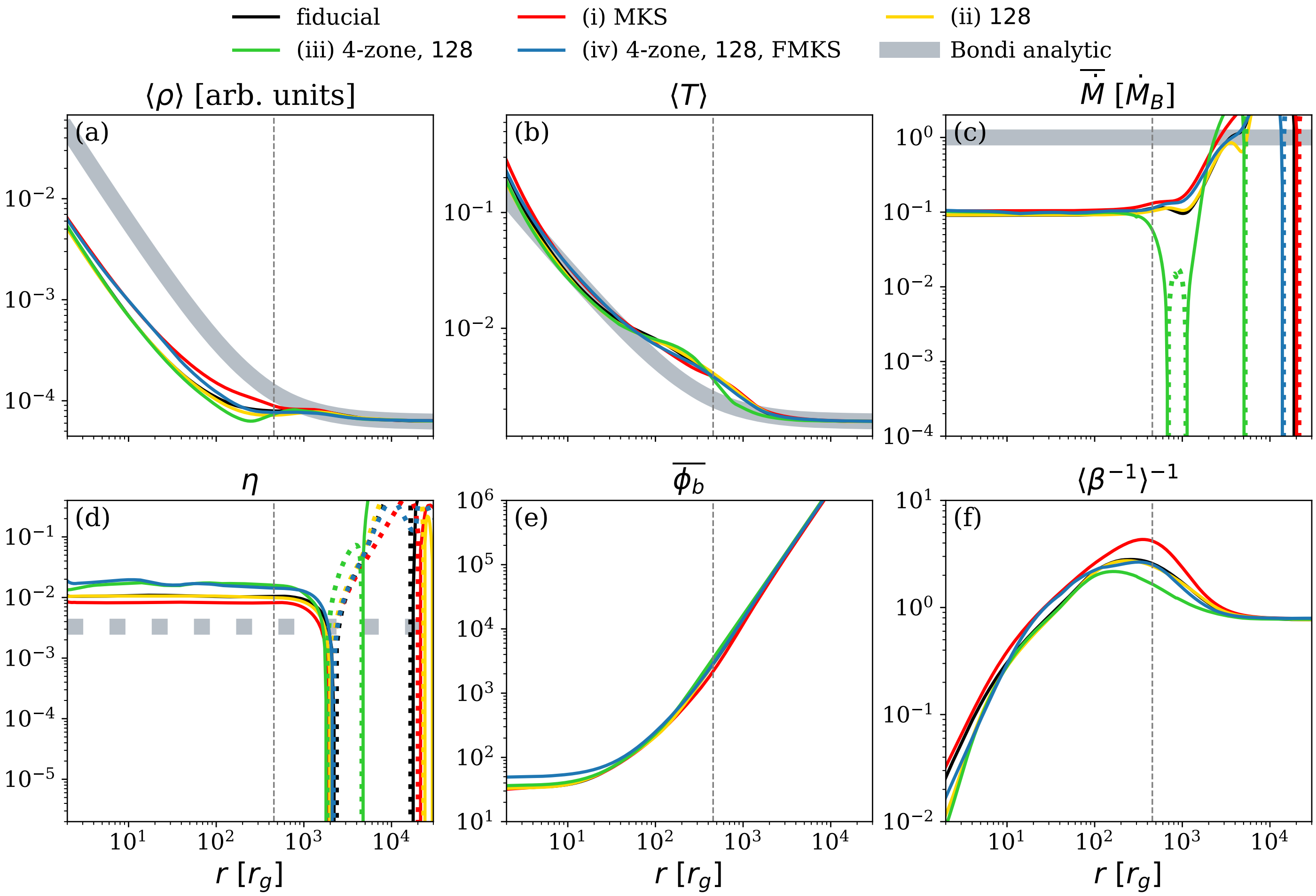

**Figure 13.** Comparison between simulations with varying resolutions and coordinate systems, some run with a single zone (one-zone) or others with the multizone method (four-zone). The run details are summarized in Table 2. The agreement across the various runs is excellent.

In the run with a factor of 0.01, the previous active zone's boundary effects are not fully dissipated away and the next active annulus can inherit that perturbed region as its new Dirichlet boundary. Through several V-cycles, any strong magnetic tension that has been generated will be preserved and even potentially be amplified by the Dirichlet boundaries, which eventually can contribute to a higher feedback efficiency.

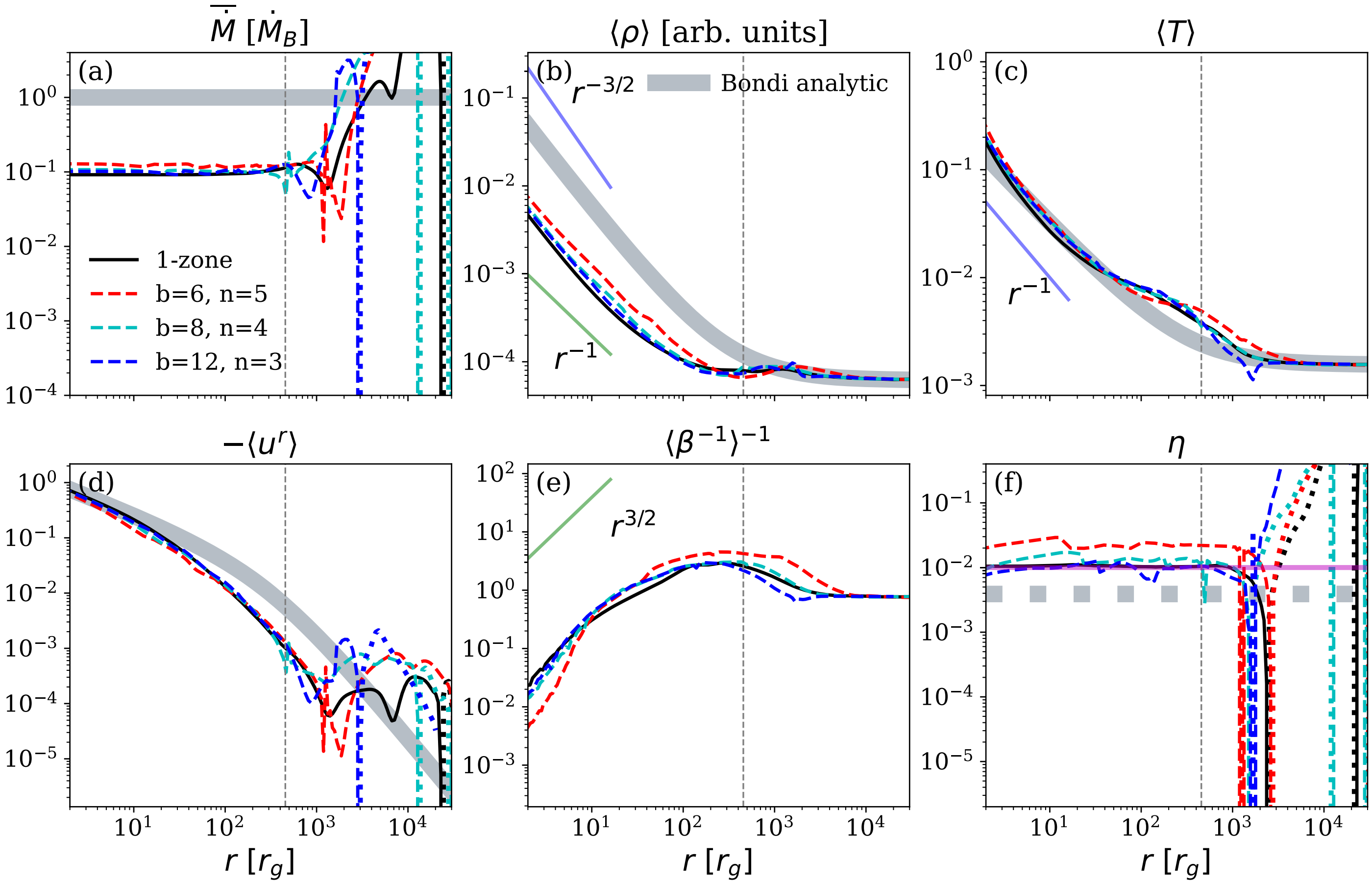

**Figure 14.** Comparison between multizone simulations with varying choices of base $b$ (colored dashed lines) and a fiducial one-zone simulation (black solid line). Note that the multizone method is not sensitive to the choice of the base except for the efficiency $\eta$ where using a small base ($b = 6$) can lead to an artificial increase in feedback.

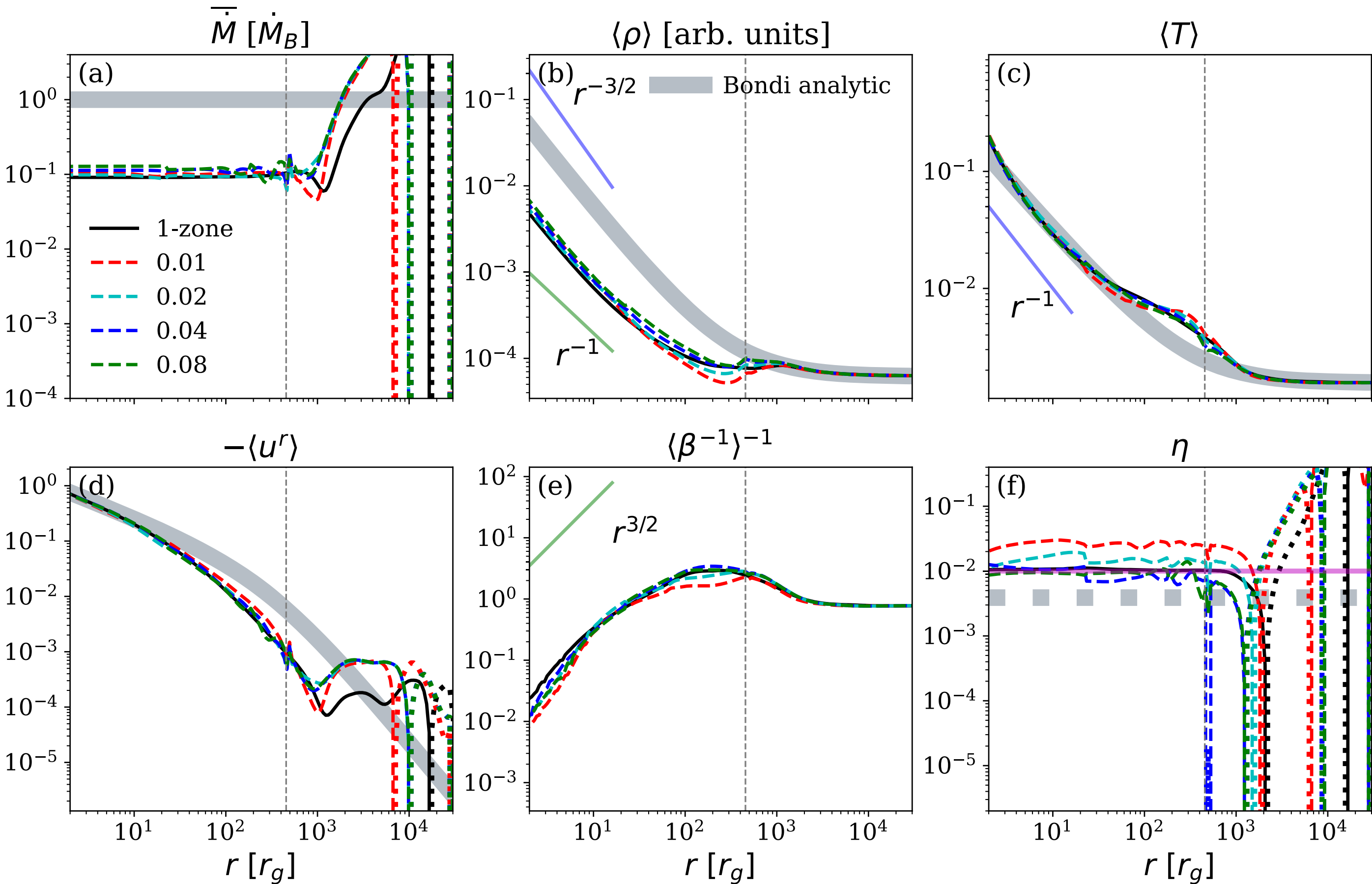

**Figure 15.** Comparison between multizone simulations with varying runtime per zone (colored dashed lines) and a fiducial one-zone simulation (black solid line). The numbers in the legend indicate the fraction of the capped characteristic timescale $t_{\mathrm{char},i}$, where the fiducial choice is 0.02 in Equation (6). When adopting a short runtime per zone with a factor of 0.01, the insufficient diffusion of the boundary effect leads to a spurious increase in feedback $\eta$. As long as the runtimes are chosen to be close to the suitable factor of 0.04, the multizone is insensitive to the choice of runtime per zone.

## Appendix G
## Time Evolution to Dynamical Equilibrium

The convergence to dynamical equilibrium is demonstrated here from the evolution of the time-averaged density and energy flux profiles. Figure 16 shows the case of the magnetized Bondi accretion simulation (`MHD` run in Section 5.4) where the time progresses from yellow to purple. After its initial relaxation phase, both the density $\rho(r)$ and efficiency $\eta(r)$ profile converge to a single profile in the last three time averages, indicative of the steady state. During the initial relaxation, the feedback efficiency $\eta(r)$ can fluctuate substantially over an order of magnitude and is not constant as a function of the radius.

Since the `MHD` run is initialized with a density profile $\rho(r)$ that resembles its final converged profile, we also present the evolution of the runs with vastly off-equilibrium initializations in Figure 17. They are initialized with a piecewise constant density and a Bondi density profile which respectively corresponds to "(v) $64^3$, const" and "(vi) $64^3$, bondi" runs in Appendix C of H. Cho et al. (2023). The simulation with an initial piecewise constant density gradually accumulates gas in Figure 17(a) because there is less gas inside the Bondi radius $R_B$ compared to its final steady state. On the other hand, the simulation starting with a Bondi density profile is initially overdense inside $R_B$ and thus undergoes a relatively violent removal of gas at early times as shown in yellow to orange time-averaged profiles in Figure 17(b). Following the sudden evacuation, the gas is gradually accreted back to find its converged state as captured by the evolving time averages. These two extra simulations have a dip in the density at around the Bondi radius $R_B$ because of a capped runtime $t_{\mathrm{MHDcap},i}$ used, an effect that is discussed in Section 5.4 and Figure 6.

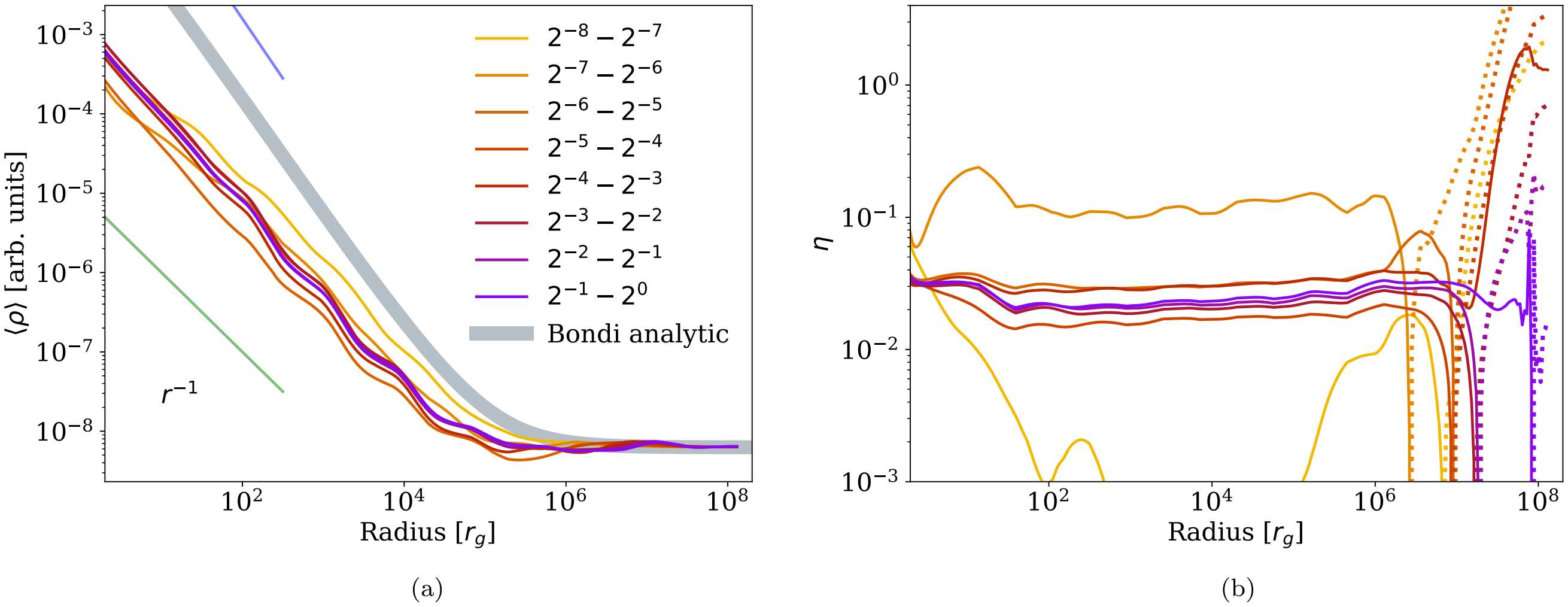

**Figure 16.** The time evolution of (a) the density $\rho$ and (b) the feedback efficiency $\eta$ profiles for the MHD run in Section 5.4. The profiles are averaged over the time range indicated in the legend where it is in units of the total runtime of the simulation. The time increases from yellow to purple.

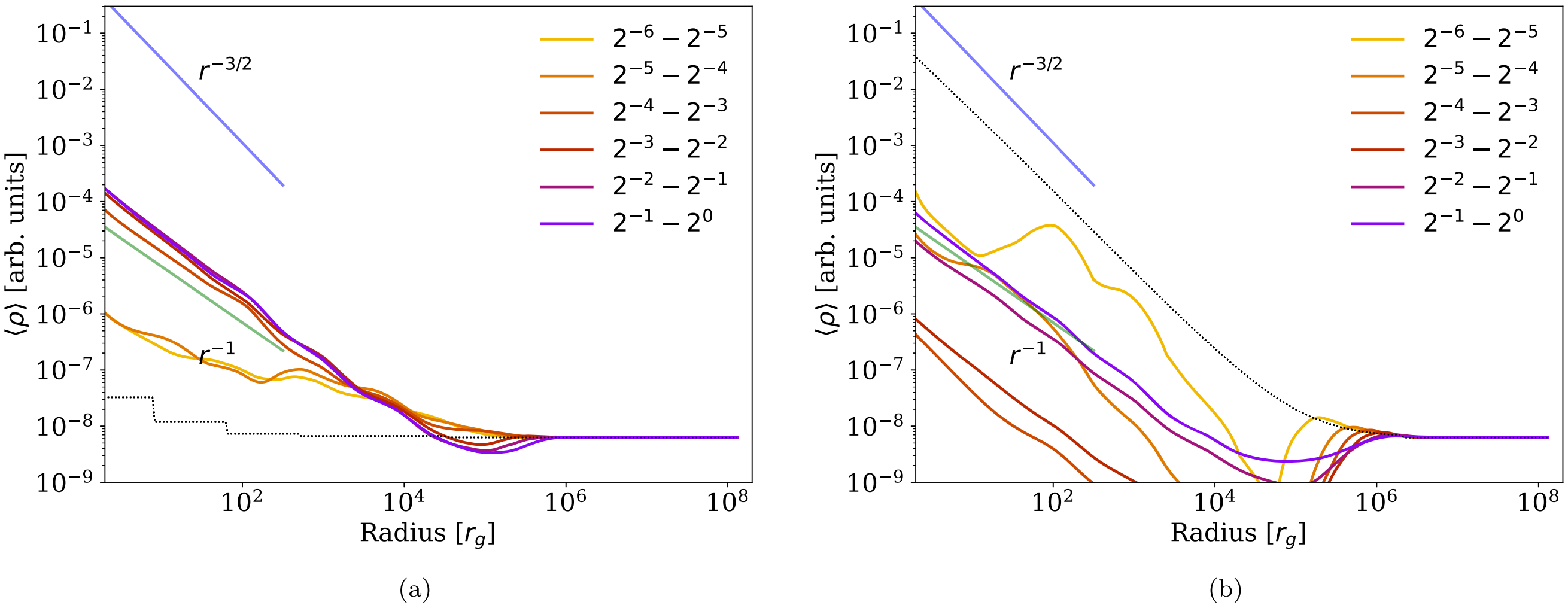

**Figure 17.** Time evolution of the density profiles for simulations with different initial densities. The black dotted lines show the initial density profiles for each simulation with (a) piecewise constant initial density and (b) Bondi initial density.

## ORCID iDs

Hyerin Cho (조혜린) https://orcid.org/0000-0002-2858-9481
Ben S. Prather https://orcid.org/0000-0002-0393-7734
Kung-Yi Su https://orcid.org/0000-0003-1598-0083
Ramesh Narayan https://orcid.org/0000-0002-1919-2730
Priyamvada Natarajan https://orcid.org/0000-0002-5554-8896